\documentclass[journal]{IEEEtran}
\usepackage{amsmath}
\usepackage{graphicx}
\usepackage{epsfig}
\usepackage{amssymb}
\usepackage{amsthm}
\usepackage{verbatim}
\usepackage{lettrine}
\usepackage{flushend,cuted}
\usepackage[top=1in, bottom=0.8in, left=0.5in, right=0.5in]{geometry}

\usepackage{url}
\usepackage{multirow}
\usepackage{cite}
\usepackage{mathrsfs}
\usepackage{cases}
\usepackage{caption}
\usepackage[english]{babel}
\usepackage[utf8x]{inputenc}
\usepackage{tikz}
\usepackage{nomencl}
\makenomenclature

\begin{document}

\title{Power System Transient Stability Assessment Using Couple Machines Method}
{\author{Songyan Wang, Jilai Yu, Wei Zhang \emph{ Member, IEEE}

\thanks{

S. Wang is with Department of Electrical Engineering, Harbin Institute of Technology, Harbin 150001, China (e-mail: wangsongyan@163.com).

J. Yu is with Department of Electrical Engineering, Harbin Institute of Technology, Harbin 150001, China (e-mail: yupwrs@hit.edu.cn).

W. Zhang is with Department of Electrical Engineering, Harbin Institute of Technology, Harbin 150001, China (e-mail: wzps@hit.edu.cn).
}
}
\maketitle
\vspace{-15pt}
\begin{abstract}

Analyzing the stability of the power system by using a few machines is promising for transient stability assessment. A hybrid direct-time-domain method that is fully based on the thinking of partial energy function is proposed in this paper. During post-fault period, a pair of machines with high rotor speed difference is defined as couple machines, and the stability analysis of the system is transformed into that of several pairs of couple machines. Based on the prediction of power-angle curve of couple machines within a sampling window after fault clearing, the proposed method avoids the definition of Center of Inertia (COI) and it can also evaluate the stability margin of the system by using the predicted power-angle curve. Simulation results demonstrate its effectiveness in transient stability assessment.

\end{abstract}


\nomenclature[1]{IMEF}{Individual machine energy function}
	
\nomenclature[2]{PEF}	{Partial energy function}
\nomenclature[3]{OMIB}	{One-machine-infinite-bus}
\nomenclature[4]{EAC}	{Equal area criterion}
\nomenclature[5]{COI}	{Center of inertia}
\nomenclature[6]{DLP}	{Dynamic liberation point}
\nomenclature[7]{DSP}	{Dynamic stationary point}
\nomenclature[8]{UEP}	{Unstable equilibrium point}
\nomenclature[9]{RUEP}	{Relevant UEP}
\nomenclature[10]{CUEP}	{Controlling UEP}
\nomenclature[11]{SVCS}	{Single-machine-and-virtual-COI-machine subsystem}
\nomenclature[12]{EEAC}	{Extended equal area criterion}
\nomenclature[13]{SIME}	{Single machine equivalence}
\nomenclature[14]{MOD}	{Mode of disturbance}
\nomenclature[15]{TSA}	{Transient stability assessment}
\nomenclature[16]{P.E.}	{Potential energy}
\nomenclature[17]{K.E.}	{Kinetic energy}
\nomenclature[18]{CCT}	{Critical fault clearing time}

\printnomenclature

\begin{center}
  \textsc{Abbreviation}
\end{center}

{PEF}\qquad{Partial energy function}

{OMIB}\quad{One-machine-infinite-bus}

{EAC}\qquad{Equal area criterion}

{COI}\qquad{Center of inertia}

{DLP}\qquad{Dynamic liberation point}

{DSP}\qquad{Dynamic stationary point}

{UEP}\qquad{Unstable equilibrium point}

{RUEP}\quad{Relevant UEP}

{CUEP}\quad{Controlling UEP}

{SVCS}\quad{Single-machine-and-virtual-COI-machine subsystem}

{EEAC}\quad{Extended equal area criterion}

{SIME}\quad{Single machine equivalence}

{MOD}\quad{Mode of disturbance}

{TSA}\qquad{Transient stability assessment}

{P.E.}\qquad{Potential energy}

{K.E.}\qquad{Kinetic energy}

{CCT}\qquad{Critical fault clearing time}

\section{Introduction}

Power systems nowadays operate in a complicated state and may be subjected to stressed operating conditions, which demands great accountability of the TSA. Since computational complexity of time-domain simulation becomes intolerable with the increase of system scale \cite{Tang1994transient}, transient energy methods are always attractive to system operators for the online transient stability assessment. In regards to the analysis based on transient energy function, the RUEP method and sustained fault method have received considerable attention and also achieved early advance \cite{Athay1979A,Kakimoto1978Transient}. Based on these works, Fouad and Stanton used a series of simulations of a practical power system to illustrate the instability phenomenon of the power system, and their work merited two conjectures \cite{Fouad1981Transient}: (i) Not all the excess kinetic energy at fault clearing contributes directly to the separation of the critical machines form the rest of the system, part of kinetic energy should be corrected.(ii) If more than one machine tends to lose synchronism, the instability is determined by the motion of some unstable critical machines.

Both conjectures marked a milestone to the transient stability studies because they originated two different direct methods. Stimulated by (i), Xue and Pavella developed the EEAC and SIME that were based on kinetic energy correction and group separation \cite{Xue1989Extended,Yin2011An,Pavella2000Transient}. EEAC and SIME are both proved to be highly efficient, and they had achieved great success in academic research as well as industrial applications. In the meaning time, enlightened by (ii), some work attempted to observe the stability of the system from a non-global perspective. In \cite{Vittal1982Power,Michel1982Power}, Vittal and Fouad stated that the instability of a single machine would be identical to the instability of the system and IMEF was firstly proposed therein. Later, Stanton performed a detailed machine by machine analysis of the energy of a multi-machine instability \cite{Stanton1982Assessment}. Stanton also defined PEF to quantify the energy transactions that make a machine transfer from stable to unstable. Then in \cite{Stanton1989Analysis,Stanton1989Transient}, PEF is utilized to quantify the energy of a local transient control action which is in contrast with the global view. Later, Rastgoufard et al. pointed out that the instability of the system is decided by a few unstable critical machines rather than non-critical machines \cite{Rastgoufard1988Multi}. The authors also utilized EAC of critical machines to determine the transient stability of a multi-machine system. Haque proposed a strategy to compute the CCT using PEF of critical machines in \cite{Haque1995Further} whereas Ando presented a potential energy ridge which can be used to predict the single machine stability \cite{Ando1988Highly}. Recently, Lu and Yu initiated a method using a pair of machines to evaluate the stability index of the system \cite{Lu2009Using}, yet the foundation as well as the mechanism of this method has not been fully investigated.
Although non-global methods are proved to be less conservative in stability analysis \cite{Michel1982Power}, these distinctive methods are at a standstill in recent years, making them a minority and fall into disuse. If one explores deep into these methods, the key problem is that the computation of the critical energy for these non-global methods is rather complicated. To be specific, in IMEF method, the evaluation of critical energy of the critical machines is fully based on the sustained fault trajectory, which means that the computation complexity of IMEF is the same as that of the sustained fault method. In PEF method, the virtual linear trajectory is applied to compute the deceleration area from fault clearing point to the DLP, however its computation is based on the time-consuming simulation of the actual simulated trajectory \cite{Stanton1989Analysis,Stanton1989Transient}.
In this paper a hybrid direct-time-domain method is proposed. This method can be seen as a continuity of the PEF method. We first clarify that the motion of a critical machine in COI reference in IMEF method and PEF method is identical to that of a two-machine subsystem consisting of a critical machine and a virtual COI machine. Then we replace the virtual COI machine in PEF with a real machine in the system, and define couple machines as the pair of real machines with high rotor-speed difference after fault clearing. Similar to the PEF method that only focuses on the stability state of the critical machine in COI reference, the proposed method only targets on analyzing the stability state of couple machines, neglecting other pairs. We prove that EAC strictly holds for couple machines and the power-angel curve (also known as Kimbark curve \cite{Stanton1989Analysis,Stanton1989Transient} ) of couple machines has a distinctive quasi-sinusoidal feature. In light of this feature, we aim to adopt EAC to assess stability index of couple machines based on the prediction of the Kimbark curve of couple machines, then compare the method with time-domain simulations and other classical transient energy methods.
Contributions of this paper are summarized as follows:

(i) The motion of a critical machine in COI reference is proved to be identical to that of a two-machine system consisting of a critical machine and a virtual COI machine.

(ii) Definition of COI is avoided in the proposed method as the virtual COI machine in PEF is replaced with a real machine in the system.

(iii) Following the thinking of PEF, the method only focuses on analyzing the stability state of couple machines, neglecting other pairs.

(iv) Strict EAC for couple machines is validated.

(v) The problem regarding critical energy computation in PEF method is overcome by using the predicted Kimbark curve of couple machines.

(vi) The method does not rely on equivalence, simplification or aggregation of the whole network or any other machines during stability analysis.

The remaining paper is organized as follows. In Section II, classic PEF method is revisited and discussed. In Section III, EAC of couple machines is analyzed. In Section IV, types of actual Kimbark curves of couple machines are defined and also predictions of these types of Kimbark curves are presented. In Section V, stability measures based on the EAC of couple machines are given. In Section VI, the selection of couple machines and supplementary details are discussed, and the the procedures of the stability analysis by using the proposed method are provided. In Section VII, the stability assessment with the proposed method is compared with time-domain simulation and classical transient energy methods. Conclusions and discussions are provided in Section VIII.
Since the proposed method is genuinely derived from PEF method, clarifications about PEF are given in advance in this paper as some concepts in \cite{Stanton1989Analysis,Stanton1989Transient} are expressed in an unsystematic and tutorial form. It is noted that “first swing” in this paper corresponds to the swing of couple machines rather than the global first-swing, also only first-swing stability of couple machines is discussed in this paper.

\section{Further Investigation on PEF Method}

\subsection{SVCS in PEF Method}

Trajectory stability theory states that power system transient stability is a trajectory stability problem rather than Lyapunov problem. In other words, transient instability of the system is identical to the “separation” among machines in the system. To describe such separation, COI is designed to depict aggregated motion of all machines in the system or groups. Both the group separation method and PEF method are based on trajectory stability theory. However, the difference between them is that the group separation method states that the instability of the system is identical to the separation between the critical group and non-critical group \cite{Xue1989Extended}, while PEF method believes that the instability of the system is identical to the separation of a single machine with respect to the system, i.e. COI of the system \cite{Stanton1989Analysis,Stanton1989Transient}.
In PEF, ``a critical machine goes unstable” should be exactly expressed as “a critical machine in COI reference goes unstable”. If we take a deep insight into the depiction of the critical machine in PEF method and IMEF method, one can find that the motion of a critical machine in COI reference is identical to the relative motion of a critical machine with respect to the virtual COI machine.
In the following analysis, the mathematical model is based on the classical model \cite{Michel1982Power}. For a n-machine system with rotor angle $\delta_{i}$ and inertia constant $M_{i}$, the motion of a single machine $i$ (no matter it is a critical machine or not) in the synchronous reference is governed by differential equations:

\begin{equation}\label{Eq_Mac_Mot}
   \begin{cases}
   \dot{\delta_{i}}=\omega_{i} \\
    M_{i}\dot{\omega_{i}}=P_{mi}-P_{ei}\\
   \end{cases}
     \setlength{\abovedisplayskip}{1pt}
  \setlength{\belowdisplayskip}{1pt}
\end{equation}
where
\newline
$P_{ei}=E_{i}^{2}G_{ii}+\sum\limits_{j=1,j\neq{i}}^{n}(C_{ij}sin\theta_{ij}+D_{ij}cos\theta_{ij})$
\newline
with
\newline
$C_{ij}=E_{i}E{j}B_{ij},D_{ij}=E_{i}E_{j}G_{ij}$
\newline
$P_{mi}$ \qquad mechanical power of machine $i$ (constant)
\newline
$E_{i}$	\qquad voltage behind transient reactance of machine $i$
\newline
$B_{ij},G_{ij}$ \quad transfer susceptance (conductance) in the reduced bus admittance matrix.

Position of the COI of the system is defined by:

\begin{equation}\label{Eq_Pos_COI}
   \begin{cases}
   \dot{\delta}_{COI}=\frac{1}{M_{T}}\sum\limits_{i=1}^{n}M_{i}\delta_{i} \\
   \omega_{COI}= \frac{1}{M_{T}}\sum\limits_{i=1}^{n}M_{i}\omega_{i} \\
    P_{COI}= \sum\limits_{i=1}^{n}(P_{mi}-P_{ei}) \\
   \end{cases}
     \setlength{\abovedisplayskip}{1pt}
  \setlength{\belowdisplayskip}{1pt}
\end{equation}
where $M_{T}=\sum\limits_{i=1}^{n}M_{i}$.

From (\ref{Eq_Pos_COI}), the motion of COI is determined by:

\begin{equation}\label{Eq_Mot_COI}
   \begin{cases}
   \dot{\delta}_{COI}=\omega_{COI} \\
    M_{T}\dot{\omega}_{COI}=P_{COI}\\
   \end{cases}
     \setlength{\abovedisplayskip}{1pt}
  \setlength{\belowdisplayskip}{1pt}
\end{equation}

Eqn. (\ref{Eq_Mot_COI}) reveals that COI can also be seen as a virtual ``machine” with its own equation of motion being described as the aggregated motion of all machines in the system.
Following (\ref{Eq_Mac_Mot}) and (\ref{Eq_Mot_COI}), since machine $i$ and COI are two ``single” machines with interactions, a two-machine subsystem can be formed by using these two machines, which is defined as a SVCS, as shown in Fig. \ref{Fig_Two_Mac_COI}.

\begin{figure}
\vspace{5pt}
\captionsetup{name=Fig.,font={small},singlelinecheck=off,justification=raggedright}
\includegraphics[width=3.5in,height=2.8in,keepaspectratio]{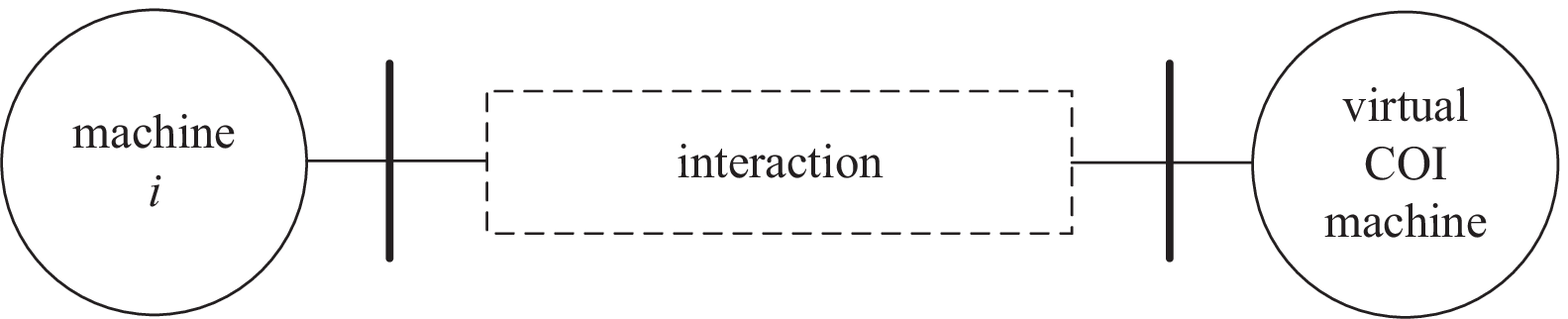}\\
  \setlength{\abovecaptionskip}{-5pt}
  \setlength{\belowcaptionskip}{0pt}
  \vspace{-2pt}
  \caption{ Two-machine subsystem formed by machine i and virtual COI machine}
  \label{Fig_Two_Mac_COI}
\end{figure}

Substituting (\ref{Eq_Mot_COI}) into (\ref{Eq_Mac_Mot}), the relative motion between the single machine and virtual COI machine of the SVCS can be given as:
\begin{equation}\label{Eq_Mot_SVC}
   \begin{cases}
   \dot{\theta}_{i}=\tilde{\omega}_{i} \\
    M_{i}\dot{\tilde{\omega}}_{i}=f_{i}\\
   \end{cases}
     \setlength{\abovedisplayskip}{1pt}
  \setlength{\belowdisplayskip}{1pt}
\end{equation}
where
\newline
$f_{i}=P_{mi}-P_{ei}-\frac{M_{i}}{M_{T}}P_{COI}$
\newline
$\theta{i}=\delta_{i}-\delta_{COI}$
\newline
$\tilde{\omega}_{i}=\omega_{i}-\omega_{COI}$

From (\ref{Eq_Mot_SVC}), one can see that the motion of each single machine (no matter it is a critical machine or a non-critical machine) in COI reference is fully identical to the relative motion of a single machine with respect to the virtual COI machine in a SVCS.

Since SVCS is a two-machine subsystem, {\it EAC strictly holds in the Kimbark curve of a critical machine in COI reference without any K.E. or P.E. corrections} \cite{Stanton1989Analysis,Stanton1989Transient,Rastgoufard1988Multi,Haque1995Further}.

\subsection{Split of the Stability of the System in PEF Method}

In transient stability analysis, critical machines are the machines that are severely disturbed by a fault and they tend to accelerate or decelerate from the rest of the machines in the system, depending on the nature of the fault. PEF analysts believe that transient behaviors of critical machines have dominant effects to the stability of the system, viz, only critical machines might go unstable and cause system to go unstable, non-critical machines are always stable and have no contribution to the instability of the system \cite{Vittal1982Power,Michel1982Power,Stanton1982Assessment,Stanton1989Analysis,Stanton1989Transient,Rastgoufard1988Multi,Haque1995Further}. Thus, PEF only focuses on analyzing the stability state of critical machines neglecting that of non-critical machines, which makes this distinctive method non-globally.
If we observe the motion of a critical machine in COI reference from the angle of SVCS, the transient stability of a multi-machine system with n machines is identical to that of a system consisting of n SVCSs. Since SVCSs with non-critical machines have no contribution to the stability of the system, the stability of the system is fully decided by SVCSs with critical machines, which implies that the stability analysis of the system in the multi-dimensional space could be fully split to low-dimensional space problems.
Stability mechanism of the PEF method and IMEF method is shown in Fig. \ref{Fig_Spl_Sta_PEF}.

\begin{figure}
\vspace{5pt}
\captionsetup{name=Fig.,font={small},singlelinecheck=off,justification=raggedright}
\includegraphics[width=3.5in,height=3.8in,keepaspectratio]{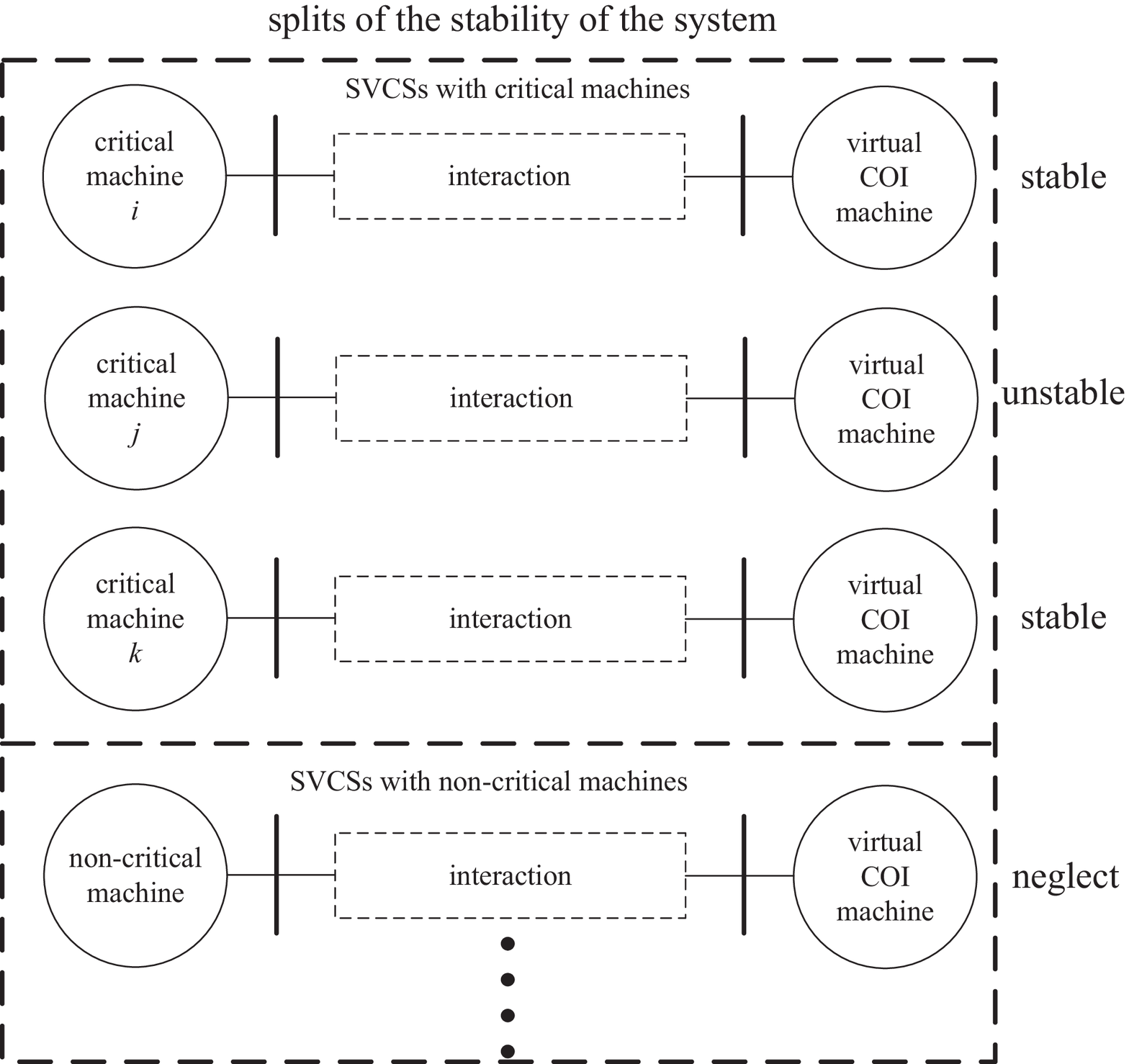}\\
  \setlength{\abovecaptionskip}{-5pt}
  \setlength{\belowcaptionskip}{0pt}
  \vspace{-2pt}
  \caption{ Two-machine subsystem formed by machine $i$ and virtual COI machine}
  \label{Fig_Spl_Sta_PEF}
\end{figure}

From Fig. \ref{Fig_Spl_Sta_PEF}, ``a critical machine in COI reference goes unstable” in PEF method and IMEF method should be further expressed as ``{\it a pair of machines consisting of a critical machine and the virtual COI machine goes unstable}”, as shown in Fig. \ref{Fig_Com_COI_Cou} (a). Although transient interactions among all machines in the system are quite complicated, the multi-machine interactions are fully implicated in the relative motion of each SVCS. Thus, the stability of the system can be judged by that of each SVCS with a critical machine independently, which is the essence of PEF method.

\begin{figure}
\vspace{5pt}
\captionsetup{name=Fig.,font={small},singlelinecheck=off,justification=raggedright}
\includegraphics[width=3.5in,height=3.8in,keepaspectratio]{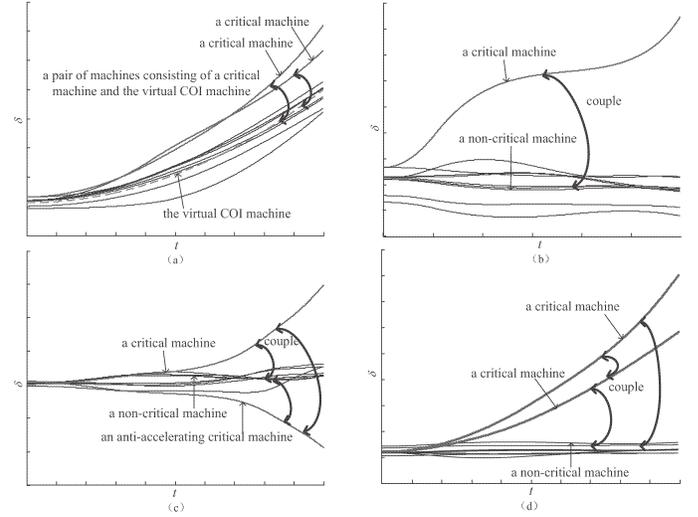}\\
  \setlength{\abovecaptionskip}{-5pt}
  \setlength{\belowcaptionskip}{0pt}
  \vspace{-2pt}
  \caption{ Similarity between a pair consisting of a critical machine and the virtual COI machine and couple machines.}
  \label{Fig_Com_COI_Cou}
\end{figure}

Following the thinking of Stanton \cite{Stanton1989Analysis,Stanton1989Transient}, transient stability principle of PEF method can be described as below:

(i) {\it The system can be judged as stable when all SVCSs with critical machines are stable}.

(ii){\it  The system can be judged as unstable as long as only one SVCS with a critical machine is found to go unstable}.

\subsection{Relative Motion of a Pair of Machines}

Assume two machines, $i$ and $j$, are brought out from the multi-machine system. A two-machine subsystem consisting of machine $i$ and machine $j$ is shown in Fig. \ref{Fig_Two_Sub_Rea}.

\begin{figure}
\captionsetup{name=Fig.,font={small},singlelinecheck=off,justification=raggedright}
  \includegraphics[width=3.5in,height=2.4in,keepaspectratio]{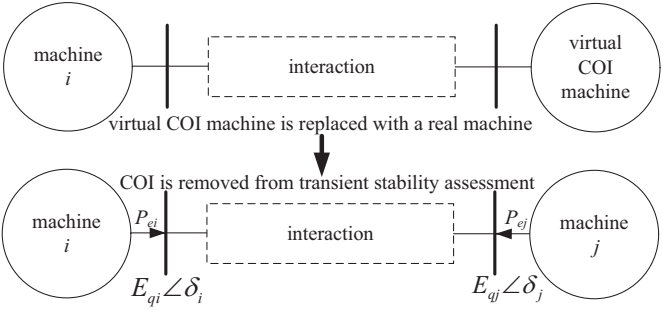}\\
  \setlength{\abovecaptionskip}{-5pt}
  \setlength{\belowcaptionskip}{0pt}
  \vspace{-2pt}
  \caption{Two-machine subsystem formed by real machines}
  \label{Fig_Two_Sub_Rea}
\end{figure}

The swing equations of the pair of machines are:
\begin{equation}\label{Eq_Swi_Cou}
   \begin{cases}
   \dot{\delta}_{i}=\omega_{i},  M_{i}\dot{\omega}_{i}=P_{mi}-P_{ei} \\
     \dot{\delta}_{j}=\omega_{j},  M_{j}\dot{\omega}_{j}=P_{mj}-P_{ej}\\
   \end{cases}
     \setlength{\abovedisplayskip}{1pt}
  \setlength{\belowdisplayskip}{1pt}
\end{equation}

Rewrite (\ref{Eq_Swi_Cou}) in an angle-difference form, we have:
\begin{equation}\label{Eq_Swi_Pai}
   \begin{cases}
   \dot{\delta}_{ij}=\omega_{ij} \\
    M_{ij}\dot{\omega}_{ij}=P_{mij}-P_{eij}\\
   \end{cases}
     \setlength{\abovedisplayskip}{1pt}
  \setlength{\belowdisplayskip}{1pt}
\end{equation}
where
\newline
$M_{ij}=\frac{M_{i}M_{j}}{M_{i}+M_{j}}$
\newline
$P_{mij}=\frac{M_{j}}{M_{i}+M_{j}}P_{mi}-\frac{M_{i}}{M_{i}+M_{n}}P_{mj}$
\newline
$P_{eij}=\frac{M_{j}}{M_{i}+M_{j}}P_{ei}-\frac{M_{i}}{M_{i}+M_{n}}P_{ej}$

Eqn. (\ref{Eq_Swi_Pai}) is the swing equation of the two machines. Similar to SVCS, machines i and j also form a two-machine subsystem. It is noted that (\ref{Eq_Swi_Pai}) merely is a mathematical transformation of (\ref{Eq_Swi_Cou}) without any simplification, aggregation or equivalence of machines or network. From the viewpoint of PEF, the swing equation of two machines underlies the stability information of the whole system during the post-fault period as both $Pe_{ij}$ and $\omega_{ij}$ are impacted by dynamic interactions of all machines in the system. In addition, all parameters in (\ref{Eq_Swi_Pai}) are defined in a synchronous reference without definition of COI.

\subsection{Strict EAC of Couple Machines}

In the following analysis, we prove that EAC strictly holds in the two real-machine subsystem along the actual system trajectory.

From (\ref{Eq_Swi_Pai}), we have:

\begin{equation}\label{Eq_EAC_Dif}
(P_{mij}-P_{eij})d\delta_{ij}= M_{ij}\omega_{ij}d\omega_{ij}
  \setlength{\abovedisplayskip}{1pt}
  \setlength{\belowdisplayskip}{1pt}
\end{equation}

Integrating the actual fault-on trajectory till fault clearing, we have

\begin{equation}\label{Eq_EAC_Int}
\int_{\delta_{ij}^{0}}^{\delta_{ij}^{c}}(P_{mij}-P_{eij}^{(F)})d\delta_{ij}= \int_{\omega_{ij}^{0}}^{\omega_{ij}^{c}}M_{ij}\omega_{ij}d\omega_{ij}
  \setlength{\abovedisplayskip}{1pt}
  \setlength{\belowdisplayskip}{1pt}
\end{equation}
where $P_{eij}^{(F)}$ corresponds to $Pe_{ij}$ during fault-on period.

Eqn. (\ref{Eq_EAC_Int}) can be further expressed as:

\begin{equation}\label{Eq_EAC_Pos}
\int_{\delta_{ij}^{0}}^{\delta_{ij}^{c}}(P_{eij}^{(F)}-P_{mij})d\delta_{ij}= \frac{1}{2}M_{ij}{\omega_{ij}^{c}}^{2}
  \setlength{\abovedisplayskip}{1pt}
  \setlength{\belowdisplayskip}{1pt}
\end{equation}

Integrating the actual post-fault trajectory after fault clearing, we have:

\begin{equation}\label{Eq_EAC_Pos_Act}
\int_{\delta_{ij^{c}}}^{\delta_{ij}}(P_{mij}-P_{eij}^{(F)})d\delta_{ij}= \int_{\omega_{ij}^{c}}^{\omega_{ij}}M_{ij}\omega_{ij}d\omega_{ij}
  \setlength{\abovedisplayskip}{1pt}
  \setlength{\belowdisplayskip}{1pt}
\end{equation}
where $P_{eij}^(PF)$  corresponds to $Pe_{ij}$ during post-fault period.

Eqn. (\ref{Eq_EAC_Pos_Act}) can be further expressed as:
\begin{equation}\label{Eq_EAC_Pos_1}
\int_{\delta_{ij}^{0}}^{\delta_{ij}}(P_{eij}^{(F)}-P_{mij})d\delta_{ij}= \frac{1}{2}M_{ij}{\omega_{ij}}^{2}- \frac{1}{2}M_{ij}{\omega_{ij}^{c}}^{2}
  \setlength{\abovedisplayskip}{1pt}
  \setlength{\belowdisplayskip}{1pt}
\end{equation}

Substituting (\ref{Eq_EAC_Pos}) into (\ref{Eq_EAC_Pos_1}) yields:
\begin{equation}\label{Eq_EAC_Fal}
\int_{\delta_{ij}^{0}}^{\delta_{ij}^{c}}(P_{mij}-P_{eij}^{(F)})d\delta_{ij}=\int_{\delta_{ij}^{c}}^{\delta_{ij}}(P_{eij}^{(F)}-P_{mij})d\delta_{ij}+ \frac{1}{2}M_{ij}{\omega_{ij}}^{2}
  \setlength{\abovedisplayskip}{1pt}
  \setlength{\belowdisplayskip}{1pt}
\end{equation}

Eqn. (\ref{Eq_EAC_Fal}) validates that EAC strictly holds in the two-machine subsystem that is formed by machine i and j, no matter these two machines are couple machines or not. Following the actual simulated system trajectory, EAC of an unstable couple machines is shown in Fig. \ref{Fig_EAC_Cou_Mac}.

\begin{figure}
\captionsetup{name=Fig.,font={small},singlelinecheck=off,justification=raggedright}
  \includegraphics[width=3.5in,height=2.4in,keepaspectratio]{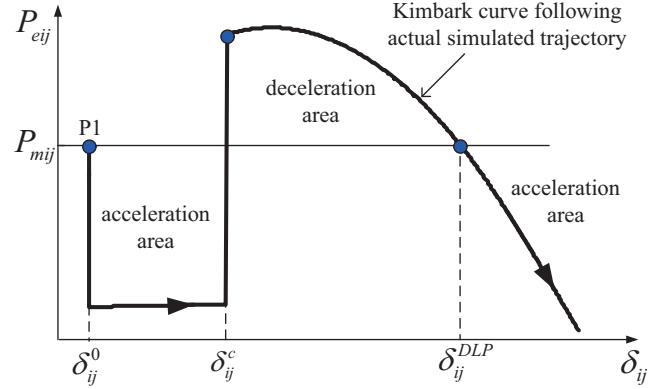}\\
  \setlength{\abovecaptionskip}{-5pt}
  \setlength{\belowcaptionskip}{0pt}
  \vspace{-2pt}
  \caption{EAC of couple machines following actual post-fault trajectory}
  \label{Fig_EAC_Cou_Mac}
\end{figure}

From Fig. \ref{Fig_EAC_Cou_Mac}, both integral parts in (\ref{Eq_EAC_Fal}) can be seen as ``areas”, which is quite similar to the EAC of the classic OMIB system. However, the EAC of couple machines is a ``precise” EAC that is based on the actual system trajectory without any K.E. or P.E. corrections, rather than the simplified EAC of the OMIB system or equivalent EAC of the EEAC system.

In following sections, EAC characteristics of couple machines are explicitly analyzed as the Kimbark curve of couple machines has a distinctive ``accelerating-decelerating” characteristic compared with that of non-couple machines. Notice that the analysis of Kimbark curves of couple machines is very simliar to that of critical machines in COI reference in the PEF method.

\subsection{Kimbark Curve of Unstable Couple Machines }

The separation of couple machines after fault is cleared can fully be reflected in the Kimbark curve. From numerous simulations, a typical Kimbark curve of unstable couple machines is shown in Fig. \ref{Fig_Kim_Cur_Uns}.

\begin{figure}
\vspace{5pt}
\captionsetup{name=Fig.,font={small},singlelinecheck=off,justification=raggedright}
  \includegraphics[width=3.5in,height=2.4in,keepaspectratio]{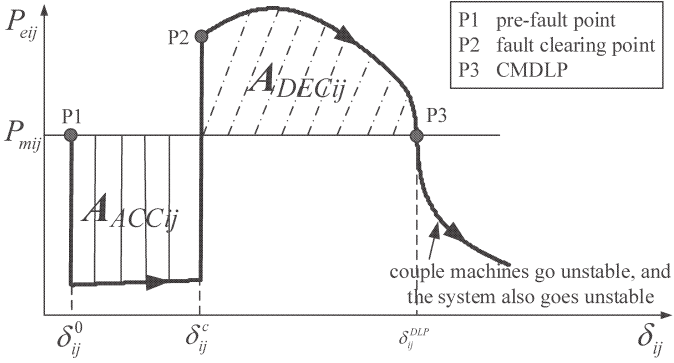}\\
  \setlength{\abovecaptionskip}{-5pt}
  \setlength{\belowcaptionskip}{0pt}
  \vspace{-2pt}
  \caption{Kimbark curve of unstable couple machines}
  \label{Fig_Kim_Cur_Uns}
\end{figure}

From Fig. \ref{Fig_Kim_Cur_Uns}, couple machines firstly accelerate from P1 to P2 during fault-on period. After fault clearing, couple machines decelerate from P2 to P3 during post-fault period. Once the system trajectory crosses P3, machine i would accelerate and separate with respect to machine j, causing couple machines go unstable. Hence, P3 can be defined as dynamic liberation point of couple machines (CMDLP) \cite{Stanton1989Transient,Stanton1989Transient}, and Peij would intersect with Pmij at CMDLP if couple machines go unstable. The Kimbark curve of unstable couple machines has a clear “accelerating-decelerating-accelerating” characteristic.

The unstable couple machines can be characterized by the occurrence of the CMDLP while velocity is positive.

\begin{equation}\label{Eq_Uns_CMDLP}
w_{ij}>0, P_{mij}-P_{eij}=0
  \setlength{\abovedisplayskip}{1pt}
  \setlength{\belowdisplayskip}{1pt}
\end{equation}

Following EAC, when couple machines go unstable, we have:

\begin{equation}\label{Eq_EAC_Uns_Mac}
A_{ACCij}>A_{DECij}
  \setlength{\abovedisplayskip}{1pt}
  \setlength{\belowdisplayskip}{1pt}
\end{equation}
where
\newline
$A_{ACCij}=\int_{\delta_{ij}^{0}}^{\delta_{ij}^{c}}(P_{mij}-P_{eij}^{(F)})d\delta_{ij}$
\newline
$A_{DECij}=\int_{\delta_{ij}^{c}}^{\delta_{ij}^{CMDLP}}(P_{eij}^{(F)-P_{mij}})d\delta_{ij}$

CMDLP is an n-dimensional point that lies in the actual post-fault system trajectory although CMDLP is described in an angle-difference form. The acceleration area, deceleration area and CMDLP of couple machines would all vary with different faults. CMDLP is used to describe first-swing instability of couple machines rather than that of the system. At $CMDLP_{ij}$, the mismatch of $Pe_{ij}$ of the couple machines formed by machine $i$ and $j$ is zero while mismatch of other pairs of machines is not zero. If more than one pair of couple machines go unstable after fault clearing, each couple would correspond to a unique CMDLP.

\subsection{Kimbark Curve of Stable or Critical Stable Couple Machines  }

When fault is not severe and couple machines are stable, the angle difference of the couple machines would initially advance after fault clearing and finally reach its maximum. In this case the velocity of the couple machines would be zero at the instant of maximum angle difference and Kimbark curve of the couple machines would not cross CMDLP. In other words, $Pe_{ij}$ would not intersect with $Pm_{ij}$. Instead, it might turn upward or turn downward because the oscillation of other machines impedes backtracking of trajectory. Under this circumstance, the couple machines would be ``stable”. A typical Kimbark curve of stable couple machines is shown in Fig. \ref{Fig_Kim_Cur_Sta}.

\begin{figure}
\vspace{5pt}
\captionsetup{name=Fig.,font={small},singlelinecheck=off,justification=raggedright}
  \includegraphics[width=3.5in,height=2.4in,keepaspectratio]{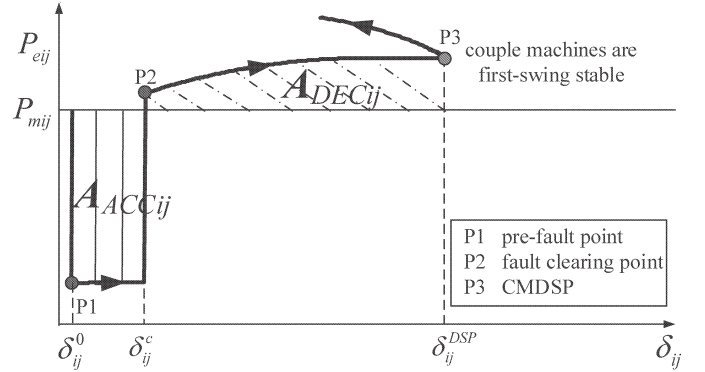}\\
  \setlength{\abovecaptionskip}{-5pt}
  \setlength{\belowcaptionskip}{0pt}
  \vspace{-2pt}
  \caption{Kimbark curve of stable couple machines}
  \label{Fig_Kim_Cur_Sta}
\end{figure}

As shown in Fig.\ref{Fig_Kim_Cur_Sta}, after fault is cleared, couple machines decelerate from P2 to P3 and the velocity of couple machines would be zero at P3, which means that couple machines is first-swing stable. Hence, P3 with zero velocity can be defined as dynamic stationary point of couple machines (CMDSP) \cite{Stanton1989Transient,Stanton1989Transient}. From the figure, it can be seen that the Kimbark curve of stable couple machines has a clear “accelerating-decelerating” characteristic.

The stable couple machines can be characterized by the occurrence of the CMDSP while deceleration power is positive.
\begin{equation}\label{Eq_Sta_CMDLP}
w_{ij}=0,P_{eij}-P_{mij}>0
  \setlength{\abovedisplayskip}{1pt}
  \setlength{\belowdisplayskip}{1pt}
\end{equation}

Following EAC, when couple machines go unstable, we have:

\begin{equation}\label{Eq_EAC_Sta_Mac}
A_{ACCij}=A_{DECij}
  \setlength{\abovedisplayskip}{1pt}
  \setlength{\belowdisplayskip}{1pt}
\end{equation}
where
\newline
$A_{DECij}=\int_{\delta_{ij}^{c}}^{\delta_{ij}^{CMDSP}}(P_{eij}^{(F)-P_{mij}})d\delta_{ij}$

In the Kimbark curve, CMDSP is the inflection point where Peij turns upward or downward. Similar to CMDLP, CMDSP is also an n-dimensional point that lies on actual post-fault system trajectory and varies with faults. CMDSP is used to describe first swing stability of a pair of couple machines rather than that of the system. At CMDSP$ij$, the velocity of the couple machines is zero while velocities of other couple machines are not zero. For more than one pair of couple machines that are stable after fault clearing, each pair of  stable couple machines would correspond to its unique CMDSP.

After fault is cleared, some couple machines might go unstable while other couples might be stable, which means that CMDSPs and CMDLPs may occur one after another along the actual post-fault system trajectory.

Following stable and unstable characterization of couple machines, the critical stable couple machines are characterized by:

\begin{equation}\label{Eq_CSta_CMDLP}
w_{ij}=0,P_{eij}-P_{mij}=0;
  \setlength{\abovedisplayskip}{1pt}
  \setlength{\belowdisplayskip}{1pt}
\end{equation}

The Kimbark curve of critical stable couple machines is shown in Fig. \ref{Fig_Kim_Cur_CSta}. It is noted that Eqn. (\ref{Eq_CSta_CMDLP}) describes an ideal critical stable state of the couple and may not hold in actual simulated cases due to the integral errors in time-domain simulations.

\begin{figure}
\vspace{5pt}
\captionsetup{name=Fig.,font={small},singlelinecheck=off,justification=raggedright}
  \includegraphics[width=3.5in,height=2.4in,keepaspectratio]{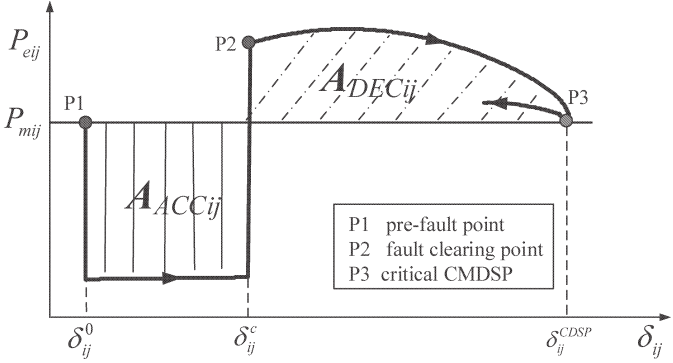}\\
  \setlength{\abovecaptionskip}{-5pt}
  \setlength{\belowcaptionskip}{0pt}
  \vspace{-2pt}
  \caption{Kimbark curve of the critical stable couple machines}
  \label{Fig_Kim_Cur_CSta}
\end{figure}

From analysis above, one can merit that the Kimbark curves of stable couple machines and unstable couple machines both have a clear “accelerating-decelerating” characteristic after fault occurs. Since the stability of the system is decided by that of couple machines, the analysis of the Kimbark curve of couple machines is of key importance in transient stability analysis. In fact, the most distinctive characteristic of the Kimbark curve of the couple machines is that it is quasi-sinusoidal and is predictable, which will be analyzed in Section IV.

\section{Predictions of the Kimbark Curve of Couple Machines}
\subsection{ Quasi-Sinusoidal Characteristic of the Kimbark Curve of Couple Machines}

Assume $\Omega_{cr}$ and $\Omega_{ncr}$ are sets of critical machines and non-critical machines after fault clearing, respectively. Machine $i$ is a critical machine that lies in $\Omega_{cr}$ and machine $j$ is a non-critical machine that lies in in $\Omega_{ncr}$, then machine $i$ and $j$ would form couple machines as their $\omega_{ij}$ is much larger than those pairs of machines brought out from the same set.
The power output of $P_{ei}$ and $P_{ej}$ is:

\begin{equation}\label{Eq_Pow_Out}
   \begin{cases}
   P_{ei}=\sum\limits_{m{\in}\Omega_{cr}}[C_{im}sin\delta_{im}+D_{im}cos\delta_{im}]\\
\qquad +\sum\limits_{n{\in}\Omega_{ncr}}[C_{in}sin\delta_{in}+D_{in}cos\delta_{in}] \\
   P_{ej}=\sum\limits_{m{\in}\Omega_{cr}}[C_{jm}sin\delta_{jm}+D_{jm}cos\delta_{jm}]\\
         \qquad +\sum\limits_{n{\in}\Omega_{ncr}}[C_{jn}sin\delta_{jn}+D_{in}cos\delta_{jn}]\\
   \end{cases}
     \setlength{\abovedisplayskip}{1pt}
  \setlength{\belowdisplayskip}{1pt}
\end{equation}

Then we have:
\begin{equation}\label{Eq_Pow_Dif}
 \begin{split}
   P_{eij} & = \frac{M_{j}}{M_{i}+M_{j}}\sum\limits_{n{\in}\Omega_{ncr}}[C_{in}sin\delta_{in}+D_{in}cos\delta_{in}]\\
 & - \frac{M_{i}}{M_{i}+M_{j}}\sum\limits_{m{\in}\Omega_{cr}}[C_{jm}sin\delta_{jm}+D_{jm}cos\delta_{jm}]\\
 & +\frac{M_{j}}{M_{i}+M_{j}}\sum\limits_{m{\in}\Omega_{cr}}[C_{im}sin\delta_{im}+D_{im}cos\delta_{im}]\\
 & - \frac{M_{i}}{M_{i}+M_{j}}\sum\limits_{n{\in}\Omega_{ncr}}[C_{jn}sin\delta_{jn}+D_{in}cos\delta_{jn}]
  \end{split}
  \setlength{\abovedisplayskip}{2pt}
  \setlength{\belowdisplayskip}{2pt}
\end{equation}

In (\ref{Eq_Pow_Dif}), $\delta_{in}$ and $\delta_{jm}$ can be expressed as:
\begin{equation}\label{Eq_Ang_Mac}
\delta_{in}=\delta_{ij}+\delta_{jn};\delta_{jm}=-\delta_{ij}+\delta_{im}
  \setlength{\abovedisplayskip}{3pt}
  \setlength{\belowdisplayskip}{3pt}
\end{equation}

In (\ref{Eq_Ang_Mac}), $\delta_{jn}$ and $\delta_{im}$ are angle differences to describe motions inside each set as $i,m\in{\Omega_{cr}}$  and  $j,n\in{\Omega_{ncr}}$.

Substituting (\ref{Eq_Ang_Mac}) into (\ref{Eq_Pow_Dif}) yields:
\begin{equation}\label{Eq_Pow_Exp}
P_{eij}=K_{sin}sin\delta_{ij}+K_{sin}\delta_{ij}+K_{tail}
  \setlength{\abovedisplayskip}{3pt}
  \setlength{\belowdisplayskip}{3pt}
\end{equation}
where
\newline
\begin{equation*}
 \begin{split}
   K_{sin} & = \frac{M_{j}}{M_{i}+M_{j}}[\sum\limits_{n{\in}\Omega_{ncr}}C_{in}cos\delta_{jn}-\sum\limits_{n{\in}\Omega_{ncr}}D_{in}sin\delta_{jn}]\\
 &  \frac{M_{i}}{M_{i}+M_{j}}[\sum\limits_{m{\in}\Omega_{cr}}C_{jm}cos\delta_{im}-\sum\limits_{m{\in}\Omega_{cr}}D_{jm}cos\delta_{im}]
  \end{split}
  \setlength{\abovedisplayskip}{2pt}
  \setlength{\belowdisplayskip}{2pt}
\end{equation*}

\begin{equation*}
 \begin{split}
   K_{cos} & = \frac{M_{j}}{M_{i}+M_{j}}[\sum\limits_{n{\in}\Omega_{ncr}}C_{in}sin\delta_{jn}+\sum\limits_{n{\in}\Omega_{ncr}}D_{in}cos\delta_{jn}]\\
 &  \frac{M_{i}}{M_{i}+M_{j}}[\sum\limits_{m{\in}\Omega_{cr}}C_{jm}sin\delta_{im}+\sum\limits_{m{\in}\Omega_{cr}}D_{jm}cos\delta_{im}]
  \end{split}
  \setlength{\abovedisplayskip}{2pt}
  \setlength{\belowdisplayskip}{2pt}
\end{equation*}

\begin{equation*}
 \begin{split}
   K_{tail} & = \frac{M_{j}}{M_{i}+M_{j}}[\sum\limits_{n{\in}\Omega_{ncr}}C_{im}sin_{im}+\sum\limits_{n{\in}\Omega_{ncr}}D_{im}cos\delta_{im}]\\
 &  \frac{M_{i}}{M_{i}+M_{j}}[\sum\limits_{m{\in}\Omega_{cr}}C_{jn}sin\delta_{jn}+\sum\limits_{m{\in}\Omega_{cr}}D_{jn}cos\delta_{jn}]
  \end{split}
  \setlength{\abovedisplayskip}{2pt}
  \setlength{\belowdisplayskip}{2pt}
\end{equation*}

Features of $K_{sin}$, $K_{cos}$ and $K_{tail}$ in (\ref{Eq_Pow_Exp}) are analyzed as follows:

(i) $K_{sin}$, $K_{cos}$ and $K_{tail}$ signify the impacts forced on the couple machines that come from all the machines in the system although $P_{eij}$ is derived from only two machines.

(ii) As majorities of post-fault trajectories of machines are still in synchronism, $\omega_{ij}$ of couple machines is much larger than  $\omega_{im}$ and  $\omega_{jn}$. Such fact implies that $\Delta\delta_{ij}$ would be much larger than $\Delta\delta_{im}$ and $\Delta\delta_{jn}$ in a short time interval, thus most components in $K_{sin}$, $K_{cos}$ and $K_{tail}$ barely change compared with violent variations of $\Delta\delta_{ij}$  during the post-fault period. In term of dynamic interactions among all machines of the system, the quasi-stationary features of $K_{sin}$, $K_{cos}$ and $K_{tail}$ are maintained due to the strong stationary effect of the majority of non-critical machines during the post-fault period. For power system transient stability analysis, a commonly used simplification for post-fault trajectory calculation is that $\delta_{im}$ and $\delta_{jn}$ are all constants. By applying this simplification, $K_{sin}$, $K_{cos}$ and $K_{tail}$ would be constant for the couple machines, thereby Kimbark curve of couple machines becomes an ideal sinusoidal curve. However, such simplification is based on the ideal assumption when fault is severe. For most actual trajectories when the fault is not severe and the system still keeps stable, the rotor motions of machines within each set would vary ($\delta_{im}$ and $\delta_{jn}$ cannot be assumed constant) irrespective of the dynamic response of machines and loads, which means that Ksin, Kcos and Ktail would be effected by the ``{\it non-sinusoidal}” feature incurred by the dynamic response of the system. Thus the shape of Kimbark curve of couple machines can be described as ``{\it quasi-sinusoidal}”.

(iii) From numerous simulations, the quasi-sinusoidal feature of Kimbark curve of couple machines still holds when the separation pattern of machines in the system is complicated.

\subsection{Parameters Identification}

According to the distinctive quasi-sinusoidal feature, Kimbark curve of couple machines can be predicted by using the parameter identification of a formula combined with sine and quadratic functions. The sine function and quadratic function are used to describe the sinusoidal and non-sinusoidal features of Kimbark curve of couple machines, respectively, and can be written as:

\begin{equation}\label{Eq_Pow_Pre}
P_{eij}^{(qr)}=(H_{q1}\delta_{ij}^{2}+H_{q2}\delta_{ij}+H_{q3})sin\delta_{ij}+H_{cos}cos\delta_{ij}+H_{cst}
  \setlength{\abovedisplayskip}{3pt}
  \setlength{\belowdisplayskip}{3pt}
\end{equation}

All parameters in (\ref{Eq_Pow_Pre}) are unknown parameters to be identified.

If $H_{q1}$ and $H_{q2}$ is set to be zero, the expression in (\ref{Eq_Pow_Pre}) would be an ideal sinusoidal curve $P_{eij}^{(sin)}$:

\begin{equation}\label{Eq_Pow_Pre_sin}
P_{eij}^{(sin)}=H_{sin}\delta_{ij}sin\delta_{ij}+H_{cos}cos\delta_{ij}+H_{cst}
  \setlength{\abovedisplayskip}{3pt}
  \setlength{\belowdisplayskip}{3pt}
\end{equation}

Parameters in (\ref{Eq_Pow_Pre}) can be identified by using the least-square approximation method, with data in a short sampling window after fault is cleared. The equation for parameter identification of $P_{eij}^{(qr)}$  is:

\begin{equation}\label{Eq_Pow_Pre_Ide}
\boldsymbol{H}_{iden}=(\boldsymbol{M}^{T}\boldsymbol{M})^{-1}\boldsymbol{M}^{T}\boldsymbol{P}_{mea}
  \setlength{\abovedisplayskip}{3pt}
  \setlength{\belowdisplayskip}{3pt}
\end{equation}
where
\newline
$\boldsymbol{H}_{iden}=[H_{q1} H_{q2} H_{q3} H_{cos} H_{cst}]$
\newline
$\boldsymbol{P}_{mea}=[P_{eij}^{(t_{1})} \cdots P_{eij}^{(t_{n})} \cdots P_{eij}^{(t_{end})} ]^{T}$
\newline
\begin{equation*}
\boldsymbol{M}\!\!=\!\!\!
 \begin{bmatrix}
     \!\!  \delta_{ij}^{2(t_{1})}sin\delta_{ij}^{(t_{1})} &  \delta_{ij}^{(t_{1})} &  sin\delta_{ij}^{(t_{1})} &  cos\delta_{ij}^{(t_{1})} & 1 \\
     \!\!\vdots &                                   \vdots           &     \vdots           &    \vdots            &  \vdots   \\
      \!\!\delta_{ij}^{2(t_{n})}sin\delta_{ij}^{(t_{n})} &  \delta{ij}^{((t_{n})} &  sin\delta_{ij}^{((t_{n})} &  cos\delta_{ij}^{((t_{n})} & 1 \\
        \!\! \vdots &                                   \vdots           &     \vdots           &    \vdots            &  \vdots  \\
 \!\!\delta_{ij}^{2(t_{end})}sin\delta_{ij}^{(t_{end})} &  \delta_{ij}^{((t_{end})} &  sin\delta{ij}^{((t_{end})} &  cos\delta_{ij}^{((t_{end})} & 1
  \end{bmatrix}
  \setlength{\abovedisplayskip}{2pt}
  \setlength{\belowdisplayskip}{2pt}
\end{equation*}

{\it \textbf{M}} in (\ref{Eq_Pow_Pre_Ide}) should be at least five dimensions as $\boldsymbol{H}_{iden}$ has five parameters to be identified. Generally the starting time point for sampling can be set as the fault clearing point, and 10 to 15 samples within 10 ms is appropriate for identification for both small and large scale systems (more than 100 machines). The identification of $P_{eij}^{(sin)}$  is similar to that of  $P_{eij}^{(qr)}$.

In (\ref{Eq_Pow_Pre_sin}), only the coefficient of $sin\delta_{ij}$ is defined by quadratic form. Theoretically, both $H_{cos}$ and $H_{cst}$ can also be replaced with quadratic forms. However, since the primary characteristic of Kimbark curve is sinusoidal, one quadratic form is efficient to describe non-sinusoidal feature of Kimbark curve.

Once $\boldsymbol{H}_{iden}$ are identified, the Kimbark curve being expressed with $P_{eij}^{(qr)}$ can be predicted. It is noted that all parameters for the predictions of Kimbark curve, i.e. $P_{eij}$ in $\boldsymbol{P}_{mea}$ and $\delta_{ij}$ in $\boldsymbol{M}$ are all local information of couple machines as COI is not used to describe couple machines, which means that the prediction of Kimbark curves of couple machines is independent of the information of other machines.

In actual simulation cases, although the quadratic form in (\ref{Eq_Pow_Pre}) can well catch the non-sinusoidal feature of Kimbark curve within the sampling window, the identification error of the quadratic form might be enlarged with the increase of simulation time. To ensure robustness of the identification, the effect of quadratic form in (\ref{Eq_Pow_Pre}) should be slightly weakened, thus following weighted formula is utilized.
\newline
\emph{Step 1} $(H_{q1}^{\prime}\delta_{ij}^{2}+H_{q2}^{\prime}\delta_{ij}+H_{q3}^{\prime})sin\delta_{ij}+H_{cos}^{\prime}cos\delta_{ij}+H_{tail}^{\prime}$ is adopted for identification at first by using simulated samples.
\newline
\emph{Step 2} Set $H_{q1}={\sigma}H_{q1}^{\prime}$ and $H_{q2}={\sigma}H_{q2}^{\prime}$, here $\sigma$ is a weighted factor and $\sigma<1$.
\newline
\emph{Step 3} Set $H_{q1}=$ and $H_{q2}$  as constants and re-identify $P_{eij}^{(qr)}$.$P_{eij}^{(qr)}$, then $H_{q3}$, $H_{cos}$ and $H_{cst}$ can be obtained.

\subsection{ Typical Types and Predictions of Actual Kimbark Curves of Couple Machines}

According to actual simulated Kimbark curves, if couple machines are stable during the first swing, the Kimbark curve might “bend down” or “bend up” in the first swing. After that, the curve might turn upward or turn downward during the second swing, as shown in Fig. \ref{Fig_Typ_Kim_Cur}.

\begin{figure}
\vspace{5pt}
\captionsetup{name=Fig.,font={small},singlelinecheck=off,justification=raggedright}
  \includegraphics[width=3.5in,height=2.4in,keepaspectratio]{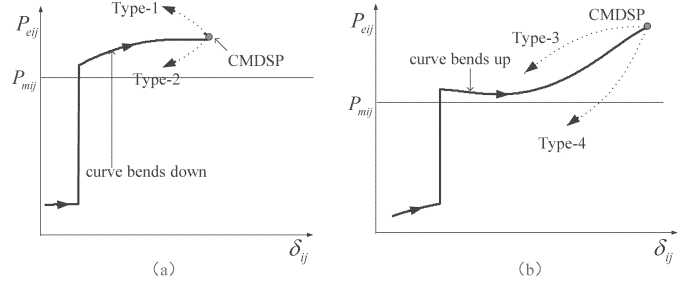}\\
  \setlength{\abovecaptionskip}{-5pt}
  \setlength{\belowcaptionskip}{0pt}
  \vspace{-2pt}
  \caption{Types of actual Kimbark curves of couple machines}
  \label{Fig_Typ_Kim_Cur}
\end{figure}

e fault clearing time the corresponding Kimbark curves can be classified into four types as shown in Fig. \ref{Fig_Typ_Kim_Cur}. The detailed illustrations are shown in Figs. \ref{Fig_Typ_One_Cur}-\ref{Fig_Typ_Fou_Cur}. The actual simulated Kimbark curves are drawn with solid lines with arrows directing actual power trajectories.

\begin{figure}
\vspace{5pt}
\captionsetup{name=Fig.,font={small},singlelinecheck=off,justification=raggedright}
  \includegraphics[width=3.5in,height=2.8in,keepaspectratio]{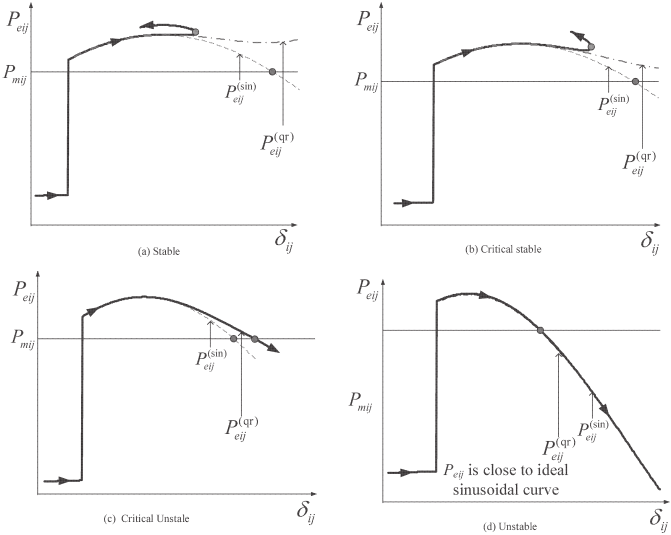}\\
  \setlength{\abovecaptionskip}{-5pt}
  \setlength{\belowcaptionskip}{0pt}
  \vspace{-2pt}
  \caption{Type-1 Curves}
  \label{Fig_Typ_One_Cur}
\end{figure}
\begin{figure}
\vspace{5pt}
\captionsetup{name=Fig.,font={small},singlelinecheck=off,justification=raggedright}
  \includegraphics[width=3.5in,height=2.8in,keepaspectratio]{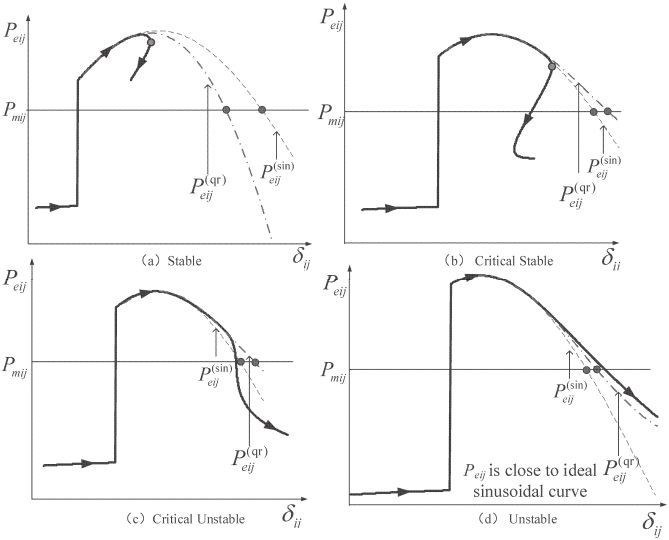}\\
  \setlength{\abovecaptionskip}{-5pt}
  \setlength{\belowcaptionskip}{0pt}
  \vspace{-2pt}
  \caption{Type-2 Curves}
  \label{Fig_Typ_Two_Cur}
\end{figure}
\begin{figure}
\vspace{5pt}
\captionsetup{name=Fig.,font={small},singlelinecheck=off,justification=raggedright}
  \includegraphics[width=3.5in,height=2.8in,keepaspectratio]{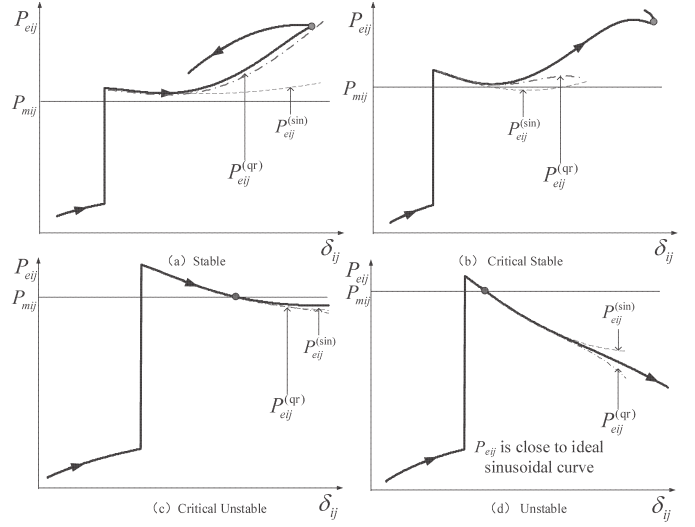}\\
  \setlength{\abovecaptionskip}{-5pt}
  \setlength{\belowcaptionskip}{0pt}
  \vspace{-2pt}
  \caption{Type-3 Curves}
  \label{Fig_Typ_Thr_Cur}
\end{figure}
\begin{figure}
\vspace{5pt}
\captionsetup{name=Fig.,font={small},singlelinecheck=off,justification=raggedright}
  \includegraphics[width=3.5in,height=2.8in,keepaspectratio]{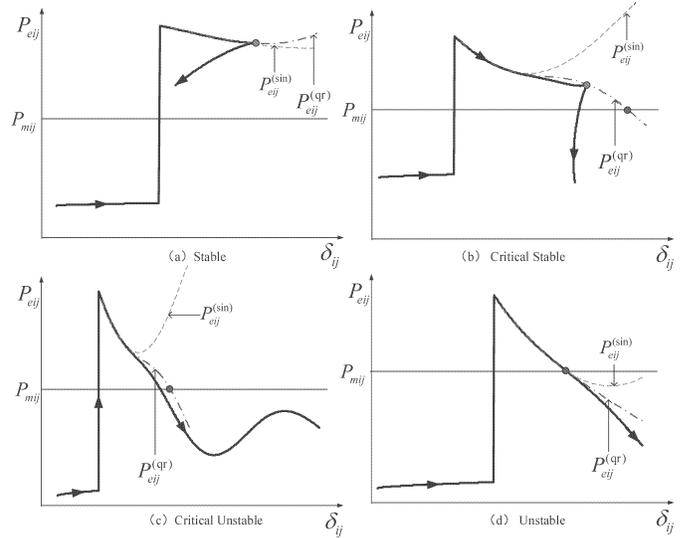}\\
  \setlength{\abovecaptionskip}{-5pt}
  \setlength{\belowcaptionskip}{0pt}
  \vspace{-2pt}
  \caption{Type-4 Curves}
  \label{Fig_Typ_Fou_Cur}
\end{figure}

From Fig. \ref{Fig_Typ_One_Cur} to Fig. \ref{Fig_Typ_Fou_Cur}, common features of all types of actual Kimbark curves are described as below.

(I) For all types of curves, the non-sinusoidal feature would be reflected when couple machines are stable.

(II) For all types of curves, the non-sinusoidal feature fades and the sinusoidal feature gradually dominates with the increase of the fault clearing time. The severer the fault, the closer Kimbark curve becomes to an ideal sinusoidal curve, which is in accord with analysis in Section IV A.

(III) All types of curves intersect with $P_{mij}$ when couple machines go unstable. The intersection point is the CMDLP as analyzed in Section III D.

(IV) All types of curves do not interest with $P_{mij}$ once the system trajectories reach CMDSP as analyzed in Section III E.

Predictions of the actual Kimbark curve by using $P_{eij}^{(qr)}$ and $P_{eij}^{(sin)}$  are also shown in Figs. \ref{Fig_Typ_One_Cur}-\ref{Fig_Typ_Fou_Cur}. Dash lines represent the predictions of $P_{eij}^{(sin)}$ while dash-dot lines are predictions of $P_{eij}^{(qr)}$. Common features of $P_{eij}^{(qr)}$  and $P_{eij}^{(sin)}$ for all types of curves are summarized as follows:

(i) Shapes of actual Kimbark curves greatly affect the approximation of $P_{eij}^{(qr)}$  and $P_{eij}^{(sin)}$.

(ii)  $P_{eij}^(qr)$ is better than $P_{eij}^(sin)$  for most stable and unstable cases. Although $P_{eij}^{(sin)}$ might be better than $P_{eij}^{(qr)}$  in a few unstable cases, the effect is slight.

(iii) The predicted $P_{eij}^{(sin)}$  and $P_{eij}^{(qr)}$  are both very close to actual Kimbark curves when couple machines go unstable because the actual curve when the fault is severe is very close to ideal sinusoidal as analyzed in Section IV A. The severer the fault, the more accurate the prediction of $P_{eij}^{(sin)}$  and $P_{eij}^{(qr)}$ is.

(iv) Comparisons between $P_{eij}^{(sin)}$  and $P_{eij}^{(sin)}$ are uncertain when couple machines are critical stable or critical unstable. However the approximation error is very slight.

From simulations above,   $P_{eij}^{(qr)}$ has a better approximation than $P_{eij}^{(sin)}$ in most cases. Although $P_{eij}^{(sin)}$  might be slightly better than $P_{eij}^{(sin)}$ in a few cases, the difference generally has no effect to stability assessment. Thus $P_{eij}^{(qr)}$  is more closer to the actual Kimbark curve than $P_{eij}^{(sin)}$ in most cases.

In fact, the predictability of Kimbark curve of couple machines is not a coincidence. Following the thinking of PEF, complicated interactions among all machines in $n$-dimensional space is split into that in $m$ two-dimensional spaces after fault clearing if $m$ pairs of couple machines exist after fault clearing. Thus, the complexity of the stability analysis can be greatly simplified by reducing the dimensions of problem solving, making Kimbark curve of couple machines sine-regular and predictable.

\section{Stability Measures of The Couple Machines}

\subsection{Predicted Deceleration Area of Couple Machines}

The predictability of the Kimbark curve means that the utilization of EAC of couple machines becomes feasible. Typical predictions of deceleration area of stable and unstable couple machines are shown in Figs. \ref{Fig_Pre_Dec_Are} (a) and (b), respectively.

\begin{figure}
\vspace{5pt}
\captionsetup{name=Fig.,font={small},singlelinecheck=off,justification=raggedright}
  \includegraphics[width=3.5in,height=1.4in,keepaspectratio]{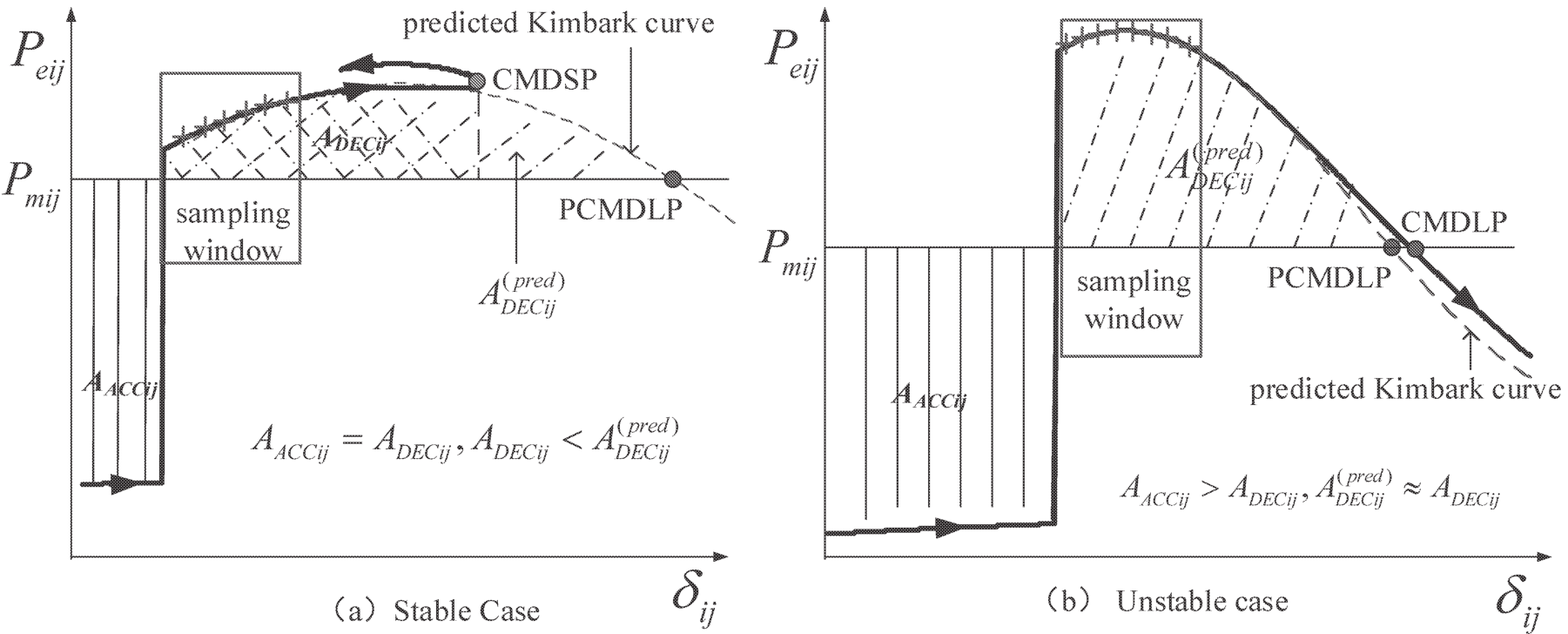}\\
  \setlength{\abovecaptionskip}{-5pt}
  \setlength{\belowcaptionskip}{0pt}
  \vspace{-2pt}
  \caption{Predicted deceleration area and PCMDLP of Couple Machines.}
  \label{Fig_Pre_Dec_Are}
\end{figure}

From Fig. \ref{Fig_Pre_Dec_Are}, the corresponding predicted CMDLP (PCMDLP) is given as the intersection point between the predicted Kimbark curve and the horizontal mechanical power curve. Thus the predicted deceleration area can be expressed as:

\begin{equation}\label{Eq_Dec_Are_Pre}
A_{ACCij}^{(Pred)}=\int_{\delta_{ij}^{c}}^{\delta_{ij}^{PCMDLP}}(P_{mij}-P_{eij}^{(qr)})d\delta_{ij}
  \setlength{\abovedisplayskip}{3pt}
  \setlength{\belowdisplayskip}{3pt}
\end{equation}
where
\newline
$A_{DECij}^{(pred)}$ predicted deceleration area
\newline
$\delta_{ij}^{PCMDLP}$ PCMDLP.

From (25), the computation of $A_{DECij}^{(pred)}$  fully relies on $P_{eij}^{(qr)}$ and PCMDLP.

Stability margin of the couple machines is defined as:
\begin{equation}\label{Eq_Sta_Mar}
\eta_{ij}=(A_{ACCij}^{(Pred)}-A_{ACCij}^{(Pred)})/A_{ACCij}
  \setlength{\abovedisplayskip}{3pt}
  \setlength{\belowdisplayskip}{3pt}
\end{equation}

In (\ref{Eq_Sta_Mar}), $\eta_{ij}>0$ means the couple is stable, $\eta_{ij}<0$ means the couple is unstable, and $\eta_{ij}=>0$ means the couple is critical stable.

The definition of $\eta_{ij}$ can be validated by Kimbark graphics. From Fig. \ref{Eq_Dec_Are_Pre} (a), if couple machines are stable, the PCMDLP would be far from CMDSP and $A_{ACCij}^{(Pred)}$ would be larger than the actual deceleration area $A_{DECij}$, which means that $\eta_{ij}$  would be positive when couple machines are stable. From Fig.\ref{Eq_Dec_Are_Pre}(b), since $P_{eij}^{(qr)}$  would almost coincide with the actual Kimbark curve when couple machines go unstable as analyzed in Section IV C, the PCMDLP would be very close to the actual one and $A_{DECij}^{(pred)}$
  would also be close to the actual $A_{DECij}$ under this circumstance. That is to say, the predicted $\eta_{ij}$  would be negative and be close to the actual stability margin when couple machines go unstable. Thus, the definition of $\eta_{ij}$  in (\ref{Eq_Sta_Mar}) is fully in accord with actual simulated stable and unstable cases of couple machines.

Since many types of actual Kimbark curves of couple machines may exist as analyzed in Section IV C, the predictions of Kimbark curves and PCMDLP of couple machines are analyzed in categories in the following sections.

\subsection{Categories of Predicted Kimbark Curves and PCMDLP}

According to variations of intersections among $P_{mij}$,  $P_{eij}^{(qr)}$ and   $P_{eij}^{(sin)}$ in Section IV, the predictions of Kimbark curves are categorized into four types as follows.

A-1:  $P_{eij}^{(qr)}$ and  $P_{eij}^{(sin)}$ both intersect with $P_{mij}$ (Fig. \ref{Fig_Typ_One_Cur}(c, d), Fig. \ref{Fig_Typ_Two_Cur}(a-d), Fig. \ref{Fig_Typ_Thr_Cur}(c, d), Fig. \ref{Fig_Typ_Fou_Cur}(d)).

A-2: Only $P_{eij}^{(qr)}$  intersects with $P_{mij}$ (Fig. \ref{Fig_Typ_Fou_Cur}(b,c)).

A-3: Only $P_{eij}^{(sin)}$  intersects with $P_{mij}$ (Fig. 10(a)-(b), Fig. \ref{Fig_Typ_Thr_Cur}(b)).

A-4: Neither  $P_{eij}^{(qr)}$ nor $P_{eij}^{(sin)}$ intersects with $P_{mij}$ (Fig. \ref{Fig_Typ_Thr_Cur}(a), Fig.\ref{Fig_Typ_Fou_Cur}(a)).

Following different types of predicted Kimbark curves, the PCMDLP of each type is defined as follows.

PCMDLP of A-1: PCMDLP is defined as the intersection point between $P_{mij}$ and  $P_{eij}^{(qr)}$   because  $P_{eij}^{(qr)}$  has a better approximation than  $P_{eij}^{(sin)}$.

PCMDLP of A-2: PCMDLP is defined as the only one intersection point between$P_{mij}$ and   $P_{eij}^{(qr)}$.

PCMDLP of A-3: PCMDLP is defined as the only one intersection point between $P_{mij}$ and  $P_{eij}^{(sin)}$.

PCMDLP of A-4: PCMDLP does not exist.
\subsection{Stability Measure of Couple Machines}
\subsubsection{ Stability Measures of A-1 and A-2}
From the Kimbark curve as in Fig. \ref{Fig_Pre_Dec_Are}, $\eta_{ij}$ can be computed by using a simple procedure as follows:

{\it Step 1} Compute $A_{ACC_{ij}}$ by the trapezoidal integration.

{\it Step 1} Divide $\delta_{ij}{\in}[\delta_{ij}^{c},\pi]$ into $M$ intervals.

{\it Step 3} Scan $\delta_{ij}$ step by step. If $P_{eii}^{(qr)}(\delta_{ii(n_{u})})-P_{mii}>0$ and $P_{eii}^{(qr)}(\delta_{ii(n_{u}+1)})-P_{mii}<0$  hold at point $n_u$, then PCMDLP is set as $(\delta_{ii(n_{u})}+\delta_{ii(n_{u}+1)})/2$ .

{\it Step 4} Compute $A_{DECij}^{(pred)}$  by the trapezoidal integration:

\begin{equation}\label{Eq_Dec_Tra}
A_{DECij}^{(pred)}=\frac{1}{2}\sum_{n=1}^{n_u}\frac{\pi-\delta_{ij}^{c}}{M}[P_{eij}^{(qr)}(\delta_{ij(n+1)})-2P_{mij}]
\end{equation}

{\it Step 5} Evaluate $\eta_{ij}$.

The computation of  $A_{DECij}^{(pred)}$ is shown in Fig. \ref{Fig_Com_Pre_Dec}.

\begin{figure}
\vspace{5pt}
\captionsetup{name=Fig.,font={small},singlelinecheck=off,justification=raggedright}
  \includegraphics[width=3.5in,height=2.8in,keepaspectratio]{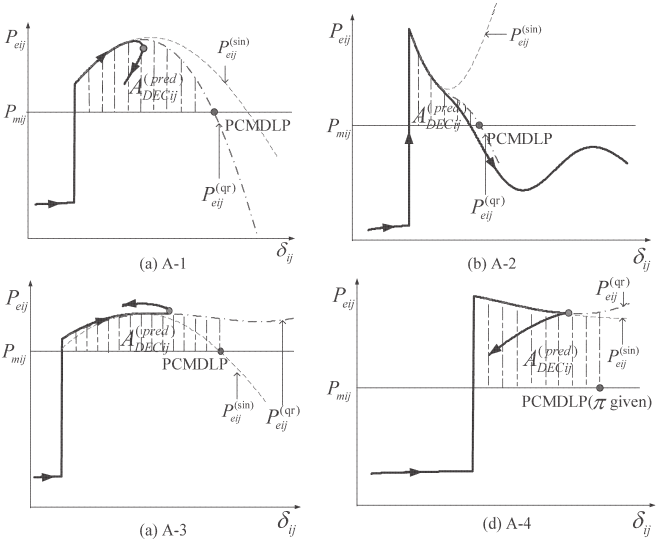}\\
  \setlength{\abovecaptionskip}{-5pt}
  \setlength{\belowcaptionskip}{0pt}
  \vspace{-2pt}
  \caption{Computation of the predicted deceleration area of A-1 to A-4.}
  \label{Fig_Com_Pre_Dec}
\end{figure}

\subsubsection{ Stability Measures of A-3}

Stability measure of A-3 is very similar to that of A-1, the only difference is that PCMDLP is computed by using the intersection point between $P_{eij}^{(sin)}$ and $Pm_{ij}$ in this case. In fact, $P_{eij}^{(sin)}$ is only used to compute PCMDLP in A-3 in the proposed method.

\subsubsection{ Stability Measures of A-4}

Computation of stability index of A-4 is special as PCMDLP does not exist. Theoretically, the predicted deceleration area can be seen as infinity in this case and the couple machines are ``infinite” stable. However, as angle difference between two machines cannot be too large, an imperial PCMDLP should be given in advance. In this paper, PCMDLP of A-4 is set as $\pi$.

Stability measure of A-4 is defined as:

\begin{equation}\label{Eq_Sta_Mea_A4}
\eta_{ij}=max\{0,(A_{DECij}^{(pred)}-A_{ACCij})/A_{ACCij}\}
\end{equation}

In (\ref{Eq_Sta_Mea_A4}), the minimum margin of couple machines should be set as zero to prevent a negative $\eta_{ij}$ because couple machines are predicted to be stable in this case.

\subsection{Stability Measure of the System}

Multi-pairs of couple machines may exist simultaneously after fault clearing. Since stability of the system is observed by multi-couples with the proposed method, the margin of the system is also defined as a multi-dimensional vector that is formed by margins of all couples.

\begin{equation}\label{Eq_Sta_Mea_Sys}
 \boldsymbol{\eta}_{sys}=[\eta_{ij}]  ij\in{\Omega_{c}}
\end{equation}

In (\ref{Eq_Sta_Mea_Sys}), $\Omega_{c}$ is the set of all couple machines. All $\eta_{ij}>0$ means the system is stable. One or a few $\eta_{ij}=0$ while the rest of $\eta_{ij}>0$ means system is critical stable. One or more $\eta_{ij}<0$ means system goes unstable.

\section{Procedures of stability analysis using proposed method}

\subsection{Identification of Couples for a Certain Fault}

For the online TSA or emergency control, the system operators may wish to grasp all severely disturbed machines in the system, thus it would be necessary to identify all couples at fault clearing point under this circumstance, and following key characteristics of couple machines can be utilized:

(i) $\omega$ of couple machines is much higher than that of non-couples at fault clearing point.

(ii) The Kimbark curve of couple machines has a clear ``accelerating-decelerating” characteristic after fault occurs.
The identification of couple machines in a small scale system may differ from that in a large scale system. Since the small system has a very weak capability to accommodate fault disturbance, all machines in a small system might be severely disturbed by faults and accelerate simultaneously in synchronous reference. Compared with small scale system, a large scale system preserves a much stronger capability to accommodate fault disturbance. To be specific, only a few machines around fault location might be disturbed while most non-critical machines that are remote from fault location may oscillate slightly after fault occurs. Thus, the identification of couple machines in a large scale system would be simpler than that in a small scale system.

From analysis above, the identification strategies of couple machines in a small scale system and a large scale system are given separately.

{\it a) Identification of couple machines in a small scale system}

{\it Step 1} Sort rotor speed of all machines in the system in a descending order at fault clearing point.

{\it Step 2} Assume the sorted index is $S=[s_{1}, s_{2} \cdots s_{N}]$. Form the set of test pairs $[s_{1}\_s_{N}, s_{2}\_s_{N-1}, \cdots s_{1+q}\_s_{N-q}]$. $q$ satisfies following:

\begin{equation}\label{Eq_Pai_Mac}
\omega_{s_{1+q+1}\_s_{N-q-1}}<\omega_{cthr},\omega_{s_{1+q}\_s_{N-q}}>\omega_{cthr}
  \setlength{\abovedisplayskip}{3pt}
  \setlength{\belowdisplayskip}{3pt}
\end{equation}

{\it Step 3} Combine first $q$ machines and last $q$ machines to form the set of candidate couple machines $\Omega_{cc}$. All combinations of candidates are $q^{2}$ pairs in $\Omega_{cc}$.

{\it Step 4} Verify candidate couples in $\Omega_{cc}$ before sampling window is over. The candidates with clear ``accelerating-decelerating” Kimbark curves are defined as couple machines and be preserved in a final set $\Omega_{c}$.

In (\ref{Eq_Pai_Mac}), $\omega_{cthr}$ is set as a threshold of selecting candidate couples. $\omega_{cthr}$ is given as 0.002 p.u.-0.005 p.u. for small scale systems based on the experiences of numerous simulations.

{\it b) Identification of couple machines in a large scale system}

{\it Step 1}  Select machines that are close to the fault location and form the set of severely disturbed machines $\omega_{se}$ at fault clearing point. Machines in $\Omega_{se}$ satisfy:
\begin{equation}\label{Eq_Pai_Mac_Lar}
\omega_{i}>\omega_{sethr} \quad i{\in}\Omega_{sr}
  \setlength{\abovedisplayskip}{3pt}
  \setlength{\belowdisplayskip}{3pt}
\end{equation}

{\it Step 2} Select a machine that is remote from the fault location as machine $j$. $\omega_{j}$ should be close to zero. Machine $j$ can be seen as the representative machine of all stationary non-critical machines in the system.

{\it Step 3} Combine machine $j$ with machines in $\Omega_{se}$ to form the set of candidate couples $\Omega_{cc}$ at fault clearing point.

{\it Step 4} The candidates with clear ``accelerating-decelerating” Kimbark curves in $\Omega_{cc}$  are defined as couple machines and are preserved in $\Omega_{c}$  after sampling window is over.

In (\ref{Eq_Pai_Mac_Lar}), $\omega_{sethr}$ is given as 0.003-0.005 p.u. for large scale systems.

\subsection{ Identification of Couples for Computation of CCT}

The identification of couple machines when computing CCT is different from that for a certain fault. The reason is that system operators do not care about all couples under this circumstance but only target the identification of critical stability of the system. Following the analysis in Section III A, the critical stability of the system is identical to the case that the lead couple is critical stable while other couples are marginal stable, thus the system operators can only focus on the lead couple neglecting other stable couples with large margins. In actual simulations, system operators may observe those a few most severely disturbed couples to ensure that the lead couple is involved. Under this circumstance, $\Omega_{c}$ can be formed with only a few elements, which greatly simplifies the identification of couples when computing CCT.

\subsection{General Procedures of the Proposed method}

General procedures by using the proposed method in TSA are outlined as below.

{\it Step 1} Select couple machines and form $\Omega_{c}$ after fault clearing.

{\it Step 2} Kimbark curve of each couple machines is predicted by using $P_{eij}^{(qr)}$ and $P_{eij}^{(sin)}$ simultaneously in a short sampling window.

{\it Step 3} For each pair of couple-machines in $\Omega_{c}$, compute $A_{ACCij}$ and $A_{DECij}$.

{\it Step 4} $\eta_{ij}$ of each couple machines is calculated using EAC of couple machines.

{\it Step 5} $\eta_{sys}$ is calculated with $\eta_{ij}$ of all couple machines, then the stability of the system is determined.

\section{Case Studies}

\subsection{ Test Network}

The proposed method is firstly tested on a small scale system, i.e., a modified 10-unit New England system (the inertia constant of Unit 1 is modified to 200 pu from 1000 pu). Three simulation cases are described below.

Case-1: Three phase short circuit event occurs at bus 34 at 0.00s and is cleared at 0.23s (the system goes unstable).

Case-2: Three phase short circuit event occurs at bus 4 at 0.00s and is cleared at 0.50s (the system goes unstable).

Case-3: Three phase short circuit event occurs at bus 4 at 0.00s and is cleared at 0.03s (the system is well stable).

Swing curves of machines in Case-1 and Case-2 are shown in Fig. 16. All parameters are set in synchronous reference.

\begin{figure}
\vspace{5pt}
\captionsetup{name=Fig.,font={small},singlelinecheck=off,justification=raggedright}
  \includegraphics[width=3.5in,height=2.8in,keepaspectratio]{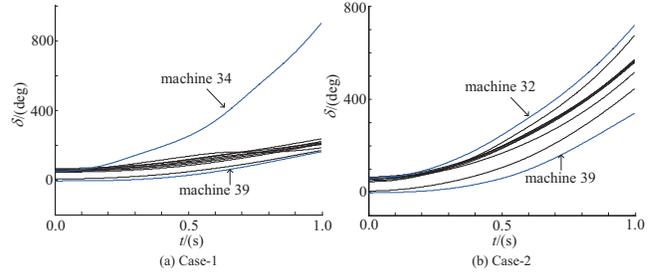}\\
  \setlength{\abovecaptionskip}{-5pt}
  \setlength{\belowcaptionskip}{0pt}
  \vspace{-2pt}
  \caption{Swing curves of machines in Case-1-Case-2.}
  \label{Fig_Swi_Cur_Mac}
\end{figure}

\subsection{Case-1}

Rotor speed of machines at fault clearing point is sorted in descending order, as shown in Table \ref{Tab_Rot_Spd}.

\begin{table}
\captionsetup{name=\textsc{Table},font={small}}
\vspace{5pt}
\centering
\setlength{\belowcaptionskip}{0pt}
\caption{\textsc{Rotor speed of all machines at fault clearing point in case-1}}
\vspace{-4pt}
\setlength{\abovecaptionskip}{2pt}
\begin{tabular}{ccc|ccc}
\hline
Order  & Items & Values(pu)& Order & Items & Values \\
\hline
1 & $\omega_{34}$        & 0.0238             & 6  &$\omega_{32}$   &0.0039        \\
2 & $\omega_{33}$        & 0.0092             & 7  &$\omega_{37}$   &0.0038         \\
3 & $\omega_{36}$       & 0.0060              & 8  &$\omega_{30}$   &0.0037          \\
4 & $\omega_{35}$       & 0.0057              & 9  &$\omega_{31}$    & 0.0033        \\
5 & $\omega_{38}$       & 0.0048              & 10 &$\omega_{39}$    & 0.0023          \\
\hline
\end{tabular}
\label{Tab_Rot_Spd}
\end{table}

According to the procedure of identification of couple machines in a small scale system as in Section VI, $q$ is set as 3. $\Omega_{cc}$ and $\Omega_{c}$ are shown in Table \ref{Tab_Ele_Set}.

\begin{table}
\captionsetup{name=\textsc{Table},font={small}}
\vspace{5pt}
\centering
\setlength{\belowcaptionskip}{0pt}
\caption{\textsc{Elements in sets in different simulated cases}}
\vspace{-4pt}
\setlength{\abovecaptionskip}{2pt}
\begin{tabular}{ccc}
\hline
Cases  & $\Omega_{cc}$ & $\Omega_{c}$ \\
\hline
\multirow{3}{*}{Case-1}  & $34\_30,34\_31,34\_39$        & $34\_30, 34\_31, 34\_39,$       \\
                         & $33\_30, 33\_31, 33\_39$,        & $33\_30, 33\_31, 33\_39,$  \\
                         & $36\_30, 36\_31, 36\_39$         & $36\_30, 36\_31, 36\_39$ \\
\hline
\multirow{3}{*}{Case-2 }  & $32\_37, 32\_30, 32\_39,$        & $32\_37, 32\_30, 32\_39, $  \\
                         & $31\_37, 31\_30, 31\_39$,        & $31\_37, 31\_30, 31\_39, $  \\
                         & $36\_37, 36\_30, 36\_39$         & $36\_37, 36\_30, 36\_39$ \\
\hline
Case-3  & $32\_39$       & $32\_39$                    \\
\hline
\multirow{3}{*}{Case-4 }  & $2\_18, 2\_19, 2\_20,$        & $\quad  $  \\
                         & $3\_18, 3\_19, 3\_20,$        & $2\_18, 2\_19, 2\_20,  $  \\
                         & $1\_18, 1\_19, 1\_20$         & $3\_18, 3\_19, 3\_20$ \\
\hline
\multirow{4}{*}{Case-6}  & XW\_LZ, HY\_LZ,        & XW\_LZ, HY\_LZ,    \\
                         & CP\_LZ, XYRD\_LZ,        & CP\_LZ, XYRD\_LZ,     \\
                         & LC1\_LZ, LC2\_LZ,          & LC1\_LZ, LC2\_LZ, \\
                         & LCR1\_LZ, LCR2\_LZ          & LCR1\_LZ, LCR2\_LZ \\
\hline
\end{tabular}
\label{Tab_Ele_Set}
\end{table}

Sampling window for approximating Kimbark curve of couples in $\Omega_c$ by using both  $P_{eij}^{(sin)}$ and $P_{eij}^{(qr)}$ is set from 230 ms to 330 ms with 10 samples (10 ms scale for each sample). Approximation types and the margin of couples are shown in Table \ref{Tab_Typ_Mar}. Kimbark curves of couples with machine 36 are very similar to those couples with machine 33 and thus are not shown here.

\begin{table}
\captionsetup{name=\textsc{Table},font={small}}
\vspace{5pt}
\centering
\setlength{\belowcaptionskip}{0pt}
\caption{\textsc{Approximation types and margin}}
\vspace{-4pt}
\setlength{\abovecaptionskip}{2pt}
\begin{tabular}{ccccc}
\hline
Cases  & Couples  &  Approximation type & Margin & Stability judgement\\
\hline
\multirow{6}{*}{Case-1}  & 34\_30	&A-1	&-0.79	&unstable     \\
                         & 34\_31	&A-1	&-0.78	&unstable \\
                         & 34\_39	&A-1	&-0.76	&unstable \\
                         & 33\_30	&A-3	&1.59	&stable \\
                         & 33\_31	&A-3	&8.65	&stable \\
                         & 33\_39	&A-3	&5.46	&stable \\
\hline
\multirow{7}{*}{Case-2 }  &32\_37	&A-1	&-0.63	&unstable  \\
                          &32\_30	&A-1	&-0.39	&unstable \\
                          & 32\_39	&A-1	&-0.47	&unstable \\
                          & 31\_37	&A-1	&-0.60	&unstable \\
                          & 31\_30	&A-1	&-0.35	&unstable \\
                          & 31\_39	&A-1	&-0.43	&unstable \\
                          & 36\_39	&A-1	&-0.18	&unstable \\
\hline
Case-3                    &32\_39   & A-1   &8.70   &table    \\
\hline
\end{tabular}
\label{Tab_Typ_Mar}
\end{table}

From Table \ref{Tab_Typ_Mar}, couple 34\_30 is the lead couple among all couples. The instability of the system is decided by unstable couples 34\_30, 34\_31 and 34\_39 because system can be judged as unstable by finding only one unstable couple. However, the severity of the system is only decided by the lead couple 34\_30 and $\eta_{sys}$ is given as -0.79.
As shown in Table II, there are total 9 pairs of couples in this case. However, if classical PEF method is used in Case-1, then only 3 critical machines, i.e., machines 34, 33 and 36 in COI reference could be identified as critical machines, thus the number of critical machines with classical PEF is smaller than the number of couples with the proposed method. The reason is that the virtual COI machine is set as the only stability reference in the classical PEF method, while 3 real machines in the system (machine 30, 31, 39) are set as stability references of couples for the proposed method, which increases the combinations of couples.

\subsection{Case-2}

To further illustrate the proposed method, Case-2 with a complicated machine-separation pattern is given. $\Omega_{cc}$ and $\Omega_{c}$ are provided in Table II. From simulation, the Kimbark curves of all couples in $\Omega_{c}$ are also very close to ideal sinusoidal although the identification of critical machines is complicated in this case. Stability judgement of unstable couple machines in Case-2 is also shown in Table \ref{Tab_Typ_Mar}. From Table \ref{Tab_Typ_Mar}, although the separation pattern of machines in Case-2 is more complicated than that of Case-1, the number of unstable couples is even larger than that of Case-1, and the system can be easily determined as unstable.

\subsection{Case-3}

Case-3 is designed to test an extreme well stable case using the proposed method. The maximum rotor speed $\omega_{32}$ at fault clearing point is only 0.0022 p.u. in this case, which indicates that system is slightly disturbed by faults. $\Omega_{c}$ with only one couple is shown in Table \ref{Tab_Ele_Set}. Margin of the couple is shown in Table \ref{Tab_Typ_Mar}. Kimbark curve of couple 32\_39 is shown in Fig. \ref{Fig_Kim_Cur_Wel_Sta}.

\begin{figure}
\vspace{5pt}
\captionsetup{name=Fig.,font={small},singlelinecheck=off,justification=raggedright}
  \includegraphics[width=3.5in,height=2.4in,keepaspectratio]{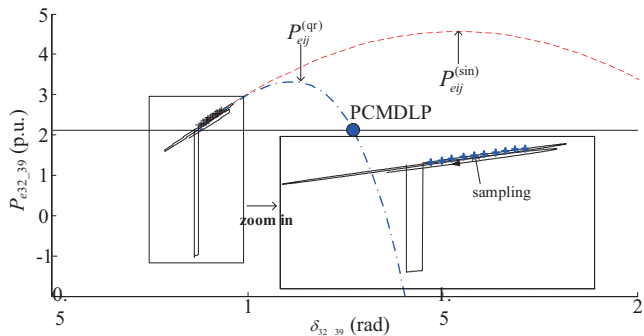}\\
  \setlength{\abovecaptionskip}{-5pt}
  \setlength{\belowcaptionskip}{0pt}
  \vspace{-2pt}
  \caption{ Kimbark curve of couple machines in a well stable case.}
  \label{Fig_Kim_Cur_Wel_Sta}
\end{figure}

From Fig. \ref{Fig_Kim_Cur_Wel_Sta}, the quasi-sinusoidal feature and “accelerating-decelerating” characteristic of actual Kimbark curve of couple machines are both preserved in the well stable case. The reason is that $\omega_{32\_39}$ is still several times larger than the rotor speed difference of non-couple machines even though $\omega_{32\_39}$ is only 0.0022 p.u in this case. Such results also validate that the proposed method can be utilized in well-stable cases effectively without over conservative judgements.

\subsection{Computation of CCT}

Different fault locations in the modified New England system are simulated to evaluate the accuracy of the proposed method, and the test results are shown in Table IV.

\begin{table}
\captionsetup{name=\textsc{Table},font={small}}
\vspace{5pt}
\centering
\setlength{\belowcaptionskip}{0pt}
\caption{\textsc{Approximation types and margin}}
\vspace{-4pt}
\setlength{\abovecaptionskip}{2pt}
\begin{tabular}{ccc}
\hline
\multirow{2}{*}{Fault location} & CCT with proposed & CCT with time domain \\
                                &  method (s)       & simulation (s) \\
\hline
bus 34	   & 0.18		&0.18 \\
bus 35		&0.29		&0.29 \\
bus 36		&0.25		&0.25 \\
bus 37		&0.21		&0.21 \\
bus 38		&0.13		&0.13 \\
bus 4		&0.45		&0.45 \\
bus 15		&0.43		&0.44 \\
bus 21		&0.32		&0.34 \\
bus 24		&0.34		&0.35 \\
\hline
\end{tabular}
\label{Tab_Cpt_CCT}
\end{table}

From Table \ref{Tab_Cpt_CCT}, the proposed method has a high accuracy in identifying CCT. The tiny error is generally incurred by approximation errors of actual Kimbark curve.

The computation of CCT at bus 34 is taken as an example to demonstrate the difference between the selection of couple machines when computing CCT and that for a certain fault in Case-1 and Case-2. With the increase of the number of iterations, it is becoming evident that couples with machine 34 are most severely disturbed couples and have leading effects to the instability of the system. Thus, as $t_c$ approaching the real CCT, the proposed method only focuses on couples 34\_30, 34\_31 and 34\_39 by neglecting other stable couples with large margins. By using the proposed method, couples 34\_30, 34\_31 and 34\_39 are all stable lead couples when $t_c$ is 0.18s and all go unstable when $t_c$ is 0.19s, thus the CCT is set as 0.18s, which fully reflects the distinctive non-global characteristic of the proposed method.

\subsubsection{Comparison between CUEP Method and Proposed Method}

Two critical unstable cases of the standard IEEE 118-bus 20-unit test system in \cite{Fouad1984Critical} are re-simulated to demonstrate the difference between the proposed method and CUEP method. The simulated critical-unstable fault clearing time is slightly different from that in \cite{Fouad1984Critical}.

Case-4: Three phase short circuit event occurs at the terminal of Gen. \#3 at 0.00s and is cleared at 0.51s.

Case-5: Three phase short circuit event occurs at the terminal of Gen. \#2 at 0.00s and is cleared at 0.20s.

{\it a) CUEP method}

For Case-4, possible MODs [Gen. \#2, Gen. \#3, Gen. \#2 and Gen. \#3] are given first, then UEP of each MOD is computed by solving a non-linear optimization problem. The MOD with Gen. \#2 is finally defined as the dominant MOD because its corresponding normalized ${\Delta}V_{PE}$ is the smallest among all possible MODs. The CUEP in the dominant MOD is ($\theta_{2,CUEP}=2.183 rad$, and global critical energy is 3.468 p.u.
For Case-5, the computation procedure is the same with that in Case-4. It is worthy to point out that both Case-4 and Case-5 have the same CUEP and same amount of global critical energy although both the fault location and fault clearing time in Case-5 are different from that in Case-4 \cite{Fouad1984Critical}.

{\it b) Proposed method}

For Case-4, $\Omega_{c}$ is given in Table II. Couples 2\_18, 2\_19 and 2\_20 go critical unstable while couples 3\_18, 3\_19 and 3\_20 are stable, then system is judged to go unstable. Kimbark curves of couples 2\_20 and 3\_20 are shown in Figs. \ref{Fig_KC_2_30}(a) and (b), respectively. Notice that the predictions of Kimbark curve of couples are both steady and robust, which technically avoids non-convergence problems of UEP computation in the CUEP method \cite{Fouad1984Critical}.

\begin{figure}
\vspace{5pt}
\captionsetup{name=Fig.,font={small},singlelinecheck=off,justification=raggedright}
  \includegraphics[width=3.5in,height=2.4in,keepaspectratio]{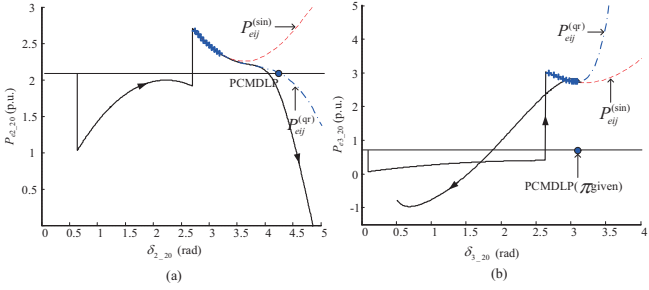}\\
  \setlength{\abovecaptionskip}{-5pt}
  \setlength{\belowcaptionskip}{0pt}
  \vspace{-2pt}
  \caption{ Kimbark curve of couple machines in a well stable case.}
  \label{Fig_KC_2_30}
\end{figure}

From Fig. \ref{Fig_KC_2_30}, Kimbark curves of couples 2\_20 and 3\_20 indicate that the critical instability of the system is fully decided by the critical instability of lead couples 2\_18, 2\_19 and 2\_20 (2\_18 and 2\_19 are not shown in the figure) rather than all couples. In fact, if the system operators only want to know system is stable or not in this case, the computation of the proposed method can be terminated immediately as long as only one couple with machine 2 is found to go unstable and analysis regarding to couples with machine 3 can be neglected. Consequently, the identification of possible MODs required in CUEP method can be avoided when judging stability of the system as it can be judged in a non-global manner by using the proposed method. However, if system operators want to obtain both stability and severity of the system, the identification of the lead couple will be needed under this circumstance and stability of all couples (couples with machine 2 and machine 3) should be judged, which becomes similar to the identification of MOD that needs  to judge stability state of all critical machines.

For Case-5, the identified couples are same with that in Case-4. However, the computed deceleration areas and PCMDLPs of couples in both cases are all different, which shows that the proposed method is more sensitive to the fault changes compared with CUEP method.

Since the proposed method is a hybrid method that relies on actual simulated Kimbark curves, the computation efficiency of the proposed might be lower than that of CUEP method in some small scale systems. However, the computation reliability of the proposed method would be revealed with the increase of system scales, as the computation of CUEP would become unbearable in large scale systems especially during online TSA.

\subsubsection{Comparison between IEEAC Method and Proposed Method}

To test the robustness of the proposed solution, three abnormal operating conditions during optimization are tested. The abnormal operating conditions include loss of communication link, disconnection of load, and loss of agent, which would produce detrimental effects if a centralized method were used.

A practical 2766-bus, 146-unit interconnected system is given to compare the difference between the proposed method and IEEAC method. SYSTEM\_LC is a regional system consisting of 8 units while SYSTM\_SD is a main system consisting of 138 units. SYSTEM\_LC and SYSTM\_SD are connected by two 500KV transmission lines. Five-order dynamic generator model with excitation and governor is utilized for simulation. The load type includes constant power load, constant impedance load, composite load and electric motor. Geographical layout of the interconnected system is shown in Fig. \ref{Fig_Pra_Sys}. Only part of SYSTEM\_SD is shown in the figure restricted by size.

\begin{figure}
\captionsetup{name=Fig.,font={small},singlelinecheck=off,justification=raggedright}
  \includegraphics[width=3.58in,height=2.82in,keepaspectratio]{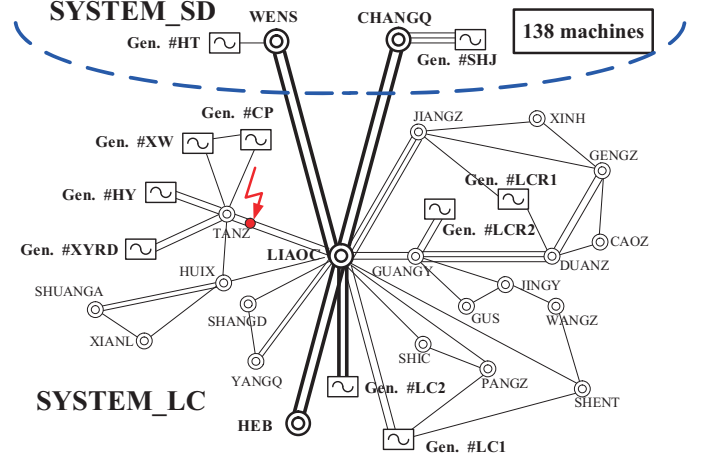}\\
    \setlength{\abovecaptionskip}{-5pt}
  \setlength{\belowcaptionskip}{0pt}
  \vspace{-5pt}
  \caption{ Kimbark curve of couples 2\_20 and 3\_20 }
  \label{Fig_Pra_Sys}
\end{figure}

The simulated case is given as below.

Case-6: The fault location is set at 90\% of Line LIAOC\_TANZ that is close to TANZ. Three phase short circuit event occurs at 0.00s and is cleared at 0.22s (CCT is 0.16s in this case).

Swing curves of machines in the interconnected system is shown in Fig. \ref{Fig_Swi_Cur_Int}.

\begin{figure}
\captionsetup{name=Fig.,font={small},singlelinecheck=off,justification=raggedright}
  \includegraphics[width=3.5in,height=2.4in,keepaspectratio]{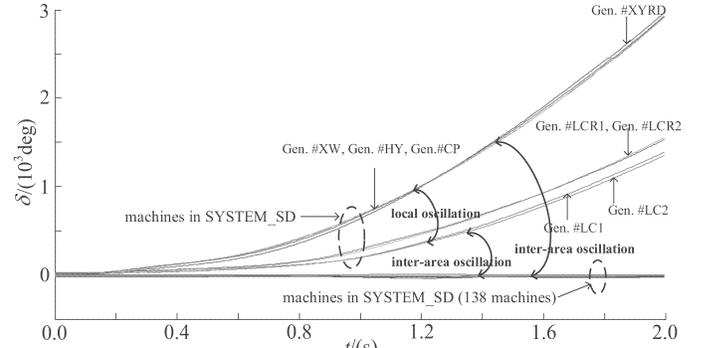}\\
  \setlength{\abovecaptionskip}{-5pt}
  \setlength{\belowcaptionskip}{0pt}
  \vspace{-5pt}
  \caption{Swing curves of machines in the interconnected system }
  \label{Fig_Swi_Cur_Int}
\end{figure}

{\it a) IEEAC method}

To simplify expression in IEEAC method,$\Omega_{n\_LC}$ is the set with [Gen. \#XW, Gen. \#HY, Gen. \#CP, Gen. \# XYRD], $\Omega_{s\_LC}$ is the set with [Gen. \#LC1, Gen. \#LC2, Gen.\#LCR1, Gen. \#LCR2], and $\Omega_{SD}$ is the set with all machines in SYSTEM\_SD. From Fig. 20, all machines in SYSTEM\_LC accelerate with respect to SYSTEM\_SD after fault is cleared, thus the separation of all machines in the whole interconnected system is a clear and favorable inter-area oscillation mode for IEEAC method.

Using IEEAC method, all machines in the interconnected system after fault clearing are separated as the critical group $\Omega_{A}$ and the non-critical group $\Omega_{S}$. From the swing curves of machines in Fig. 20, possible group separation patterns are given as below.

Pattern-1: $\Omega_{A}$ is set as $\Omega_{n\_LC}$ and $\Omega_{S}$ is set as $\Omega_{s\_LC} \cup \Omega_{SD}$.

Pattern -2: $\Omega_{A}$ is set as $\Omega_{n\_LC} \cup \Omega_{s\_LC}$  and $\Omega_{S}$ is set as $\Omega_{SD}$.

Pattern-2 is finally set as the dominated group separation pattern because stability margin of the OMIB system in Pattern-2 is much smaller than that in Pattern-1. For Pattern-2, 8 machines in $\Omega_{n\_LC}$ and $\Omega_{S}$ and 138 machines in  $\Omega_{SD}$ are aggregated as Machine-A and Machine-S, respectively. The Kimbark curve of aggregated OMIB system in Pattern-2 is shown in Fig. \ref{Fig_KC_IEEAC}. The equivalent $P_m$ in the figure is not a horizontal line as governor is deployed.

\begin{figure}
\captionsetup{name=Fig.,font={small},singlelinecheck=off,justification=raggedright}
  \includegraphics[width=3.7in,height=2.4in,keepaspectratio]{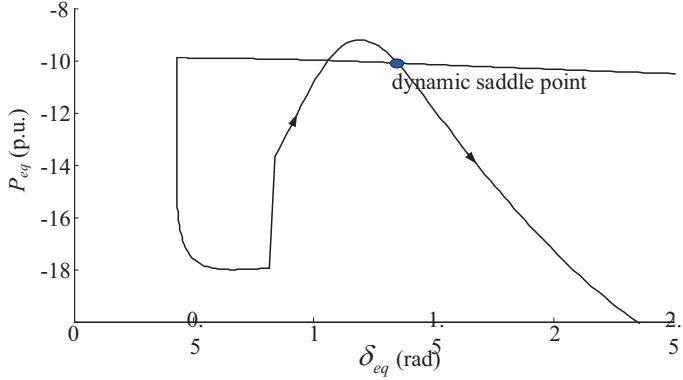}\\
  \setlength{\abovecaptionskip}{-5pt}
  \setlength{\belowcaptionskip}{0pt}
  \vspace{-5pt}
  \caption{ Kimbark curve of the OMIB system using IEEAC method}
  \label{Fig_KC_IEEAC}
\end{figure}

From Fig. \ref{Fig_KC_IEEAC}, one can see that the Kimbark curve of the OMIB system goes across dynamic saddle point [6] which lies in actual post-fault trajectory at 0.37s, thus the interconnected system is judged to go unstable and inter-area oscillation is identified.

{\it b) Proposed method}

Couple identification strategy in the large scale system is utilized first using proposed method. Targeting machines that are close to fault location, $\Omega_{se}$ is formed by [XW, HY, CP, XYRD, LC1, LC2, LCR1, LCR2]. Gen. \#LZ in SYSTEM\_SD (not shown in Fig. \ref{Fig_Pra_Sys}) that is quite remote from the fault location is selected as machine $j$. Gen. \#LZ can be seen as the representative machine of the main SYSTEM\_SD. $\Omega_{cc}$ and $\Omega_{c}$ are shown in Table \ref{Tab_Ele_Set}.

From Table \ref{Tab_Ele_Set}, unlike IEEAC method that separates all machines in the interconnected system into two groups, the proposed method only focuses on the stability of 8 couples in $\Omega_{c}$ that are formed with generators in SYSTEM\_LC and Gen. \#LZ in SYSTEM\_SD. Using proposed method, all couples in $\Omega_{c}$ are judged to go unstable, which reveals that all machines in SYSTEM\_LC separate with respect to the same machine Gen. \#LZ in SYSTEM\_SD, thus the inter-area oscillation is substantially identified. Kimbark curves of two unstable couples HY\_LZ and LCR1\_LZ in $\Omega_{c}$ are shown in Fig. \ref{Fig_KC_Int_P} (a) and (b), respectively.

\begin{figure}
\captionsetup{name=Fig.,font={small},singlelinecheck=off,justification=raggedright}
  \includegraphics[width=3.5in,height=2.4in,keepaspectratio]{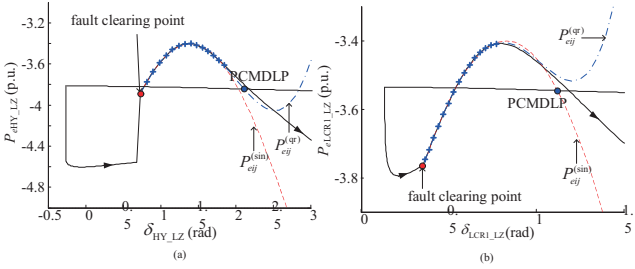}\\
  \setlength{\abovecaptionskip}{-5pt}
  \setlength{\belowcaptionskip}{0pt}
  \vspace{-5pt}
  \caption{ Kimbark curves of couples in the interconnected system }
  \label{Fig_KC_Int_P}
\end{figure}

From Fig. \ref{Fig_KC_Int_P}, the quasi-sinusoidal characteristic is also fully reflected by Kimbark curves of couples in the large scale system and is well captured by the proposed method, which validates the analysis in Section IV.

Since all machines in regional SYSTEM\_LC are severely disturbed, it is also worthy for system operators to obtain the local-oscillation situation inside SYSTEM\_LC although inter-area oscillation has already been identified. Using Gen. \#LC1 as the reference machine of SYSTEM\_LC, swing curves of all machines in regional SYSTEM\_LC are shown in Fig. \ref{Fig_SC_LC}.

\begin{figure}
\captionsetup{name=Fig.,font={small},singlelinecheck=off,justification=raggedright}
  \includegraphics[width=3.5in,height=2.4in,keepaspectratio]{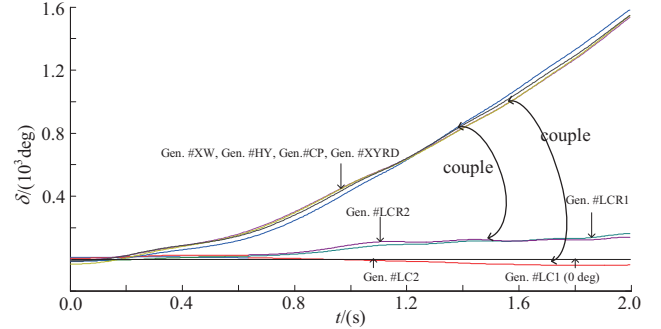}\\
  \setlength{\abovecaptionskip}{-5pt}
  \setlength{\belowcaptionskip}{0pt}
  \vspace{-5pt}
  \caption{Swing curves of machines in the regional SYSTEM\_LC }
  \label{Fig_SC_LC}
\end{figure}

From Fig. \ref{Fig_SC_LC}, local oscillation also occurs in regional SYSTEM\_LC. Using the proposed method, 16 couples are formed inside SYSTEM\_LC following the couple identification strategy in the small scale system, as shown in Fig. \ref{Fig_SC_LC}. Kimbark curves of two representative unstable couples HY\_LCR1 and HY\_LCR2 inside SYSTEM\_LC are shown in Fig. \ref{Fig_KC_LC_Re} (a) and (b), respectively.

\begin{figure}
\vspace{5pt}
\captionsetup{name=Fig.,font={small},singlelinecheck=off,justification=raggedright}
  \includegraphics[width=3.6in,height=2.4in,keepaspectratio]{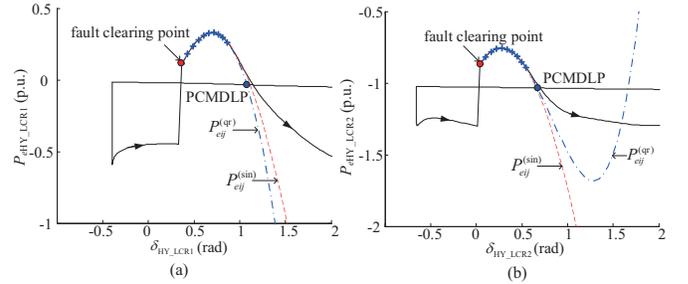}\\
  \setlength{\abovecaptionskip}{-5pt}
  \setlength{\belowcaptionskip}{0pt}
  \vspace{-5pt}
  \caption{Kimbark curves of couples in the regional SYSTEM\_LC.}
  \label{Fig_KC_LC_Re}
\end{figure}

From simulations, all couples inside SYSTEM\_LC go unstable and the local oscillation inside regional SYSTEM\_LC is also identified by the proposed method.

An extreme case is also given to further demonstrate the non-global characteristic and local-oscillation identification of the proposed method in the online TSA. Assume the angle-measurements of all machines in SYSTEM\_LC except Gen. \#HY and Gen. \#LCR1 malfunction, then the rotor angles of most machines in SYSTEM\_LC become “unobservable” for the system operators after fault occurs, leaving only Gen. \#HY and Gen. \#LCR1 “observable”.
Under such extreme circumstance, the inter-area oscillation could not be identified by the proposed method (or any other methods) as rotor angles of most machines in SYSTEM\_LC are unknown. Meanwhile, $\eta_{sys}$ is also unknown as system operators do not know whether couple HY\_LCR1 is the lead couple or not. However, the local oscillation inside SYSTEM\_LC still can be identified by the proposed method as the only “observable” couple HY\_LCR1 can be easily judged to go unstable.

{\it c) Computation efficiency analysis }

Compared with IEEAC method that can only identify inter-area oscillation in this case, both inter-area oscillation and local oscillation can be identified using the proposed method. In addition, two parallel computations are deployed simultaneously in the proposed method.

(i) Inter-area oscillation mode and local oscillation mode are identified in parallel.

(ii) In each mode, the stability of each couple is judged in parallel.

Since the selection of couples and parameters identification of Kimbark curves in both oscillation modes are simple, the corresponding time consumed can be neglected and the total time consumed is just the length of sampling window for each couple within 150 ms after fault is cleared using the proposed method.

In addition, compared with IEEAC method that relies on transient information of all machines in the interconnected system, local information of only 9 machines is used for online TSA using proposed method without any aggregation or equivalence. To be specific, the local information of each machine for TSA is $\delta_{i}$ and $P_{ei}$ from the pre-fault point to the end point of the sampling window, and $\omega_{i}^{c}$  at fault clearing point.

From simulations above, one can find that different oscillation modes require different identification strategies of couples using the proposed method. Then one question emerges: Could a couple inside SYSTEM\_LC also be defined as a couple in the interconnected system? In fact, the formation of couple machines is an intrinsic nature of the transient stability and is fully independent of definition of regional or interconnected systems. For instance, if couple HY\_LCR1 is a couple in SYSTEM\_LC, then it is also a couple in the interconnected system as SYSTEM\_LC is a regional system inside the interconnected system. That is to say, SYSTEM\_LC and the interconnected system can both be judged to go unstable once couple HY\_LCR1 goes unstable. However, for the TSA of the interconnected system, the system operator would only focus on the inter-area oscillation to grasp the whole vision of the instability of the interconnected system, thus HY\_LCR1 is theoretically a real couple but is methodologically neglected in the couple identification strategy of the interconnected system because the instability of HY\_LCR1 has no contribution to the identification of the inter-area oscillation.

{\it d) Discussion of the relationship between IEEAC method and proposed method}

IEEAC method is fully based on trajectory stability theory as the Kimbark curve of the equivalent OMIB system and the dynamic saddle point are fully obtained from actual simulated trajectory, which means that the stability margin and dynamic saddle point of the OMIB system would vary immediately with the slight change in fault location or fault clearing time. Compared with other classical global methods, IEEAC method is more sensitive to the fault change and could give a more accurate stability judgement, which is also the major advantage of IEEAC method.
Similar to the theoretical basis of IEEAC method, the proposed method is also based on trajectory stability theory and CMDLP also lies in the actual simulated post-fault trajectory. The difference between the IEEAC method and the proposed method is the angle of observing the transient instability of the system. For the proposed method, the instability of the system is described as the separation of a few pairs of couple machines, rather than separation of groups.

\section{Conclusion and discussion}

This paper proposes a hybrid direct-time-domain method. The stability of a critical machine in IMEF and PEF is clarified as the stability of a pair of machines consisting of a critical machine and the virtual COI machine. Following the thinking of PEF method, the proposed method transforms the stability analysis of the system into that of a few pairs of couple machines, which avoids the equivalence or aggregation of network. The stability index of couple machines is evaluated by EAC individually and instability of the system is determined by unstable couples, which greatly reduces complexity of stability analysis. The method developed is able to efficiently evaluate the system stability and simulation results obtained are practical in the sense that they are sufficiently accurate compared with time domain simulation and classic global methods.

Further work is to focus on fundamental theories and multi-swing stability judgement with the proposed method.

\end{document}